\documentclass[3p,authoryear,a4paper]{elsarticle}
\usepackage{graphicx} 
\usepackage{parskip} 
\usepackage{xcolor} 
\usepackage{booktabs} 

\usepackage[font={small}]{caption}
\usepackage{subfig}
\usepackage[section]{placeins} 

\usepackage{natbib} 
\usepackage[hyphens]{url} 

\usepackage{hyperref} 
\hypersetup{
    colorlinks=false,
    linktoc=all, 
    hidelinks
}

\usepackage{minted} 
\newminted{python}{
    fontsize=\footnotesize
}

\usepackage{tikz}
\usetikzlibrary{positioning, fit}
\usetikzlibrary {shapes.geometric}

\usepackage{pgfplots} 

\usepackage{amsmath} 
\usepackage{amssymb}
\usepackage{listofitems} 
\usetikzlibrary{arrows.meta} 
\usepackage[outline]{contour} 
\contourlength{1.4pt}
\colorlet{myred}{red!80!black}
\colorlet{myblue}{blue!80!black}
\colorlet{mygreen}{green!60!black}
\colorlet{myorange}{orange!70!red!60!black}
\colorlet{mydarkred}{red!30!black}
\colorlet{mydarkblue}{blue!40!black}
\colorlet{mydarkgreen}{green!30!black}

\usepackage{xcolor}
\definecolor{tolgreen}{RGB}{51, 117, 56}
\definecolor{tolgold}{RGB}{220, 205, 125}
\definecolor{tolcoral}{RGB}{194, 106, 119}

\definecolor{cbpink}{HTML}{D81B60}
\definecolor{cbblue}{HTML}{1E88E5}
\definecolor{cbyellow}{HTML}{FFC107}

\tikzset{
  >=latex, 
  node/.style={thick,circle,draw=myblue,minimum size=22,inner sep=0.5,outer sep=0.6},
  node in/.style={node,green!20!black,draw=mygreen!30!black,fill=mygreen!25},
  node hidden/.style={node,blue!20!black,draw=myblue!30!black,fill=myblue!20},
  node convol/.style={node,orange!20!black,draw=myorange!30!black,fill=myorange!20},
  node out/.style={node,red!20!black,draw=myred!30!black,fill=myred!20},
  connect/.style={thick,mydarkblue}, 
  connect arrow/.style={-{Latex[length=4,width=3.5]},thick,mydarkblue,shorten <=0.5,shorten >=1},
  node 1/.style={node in}, 
  node 2/.style={node hidden},
  node 3/.style={node out}
}

\definecolor{darkspringgreen}{rgb}{0.09, 0.45, 0.27}
\definecolor{goldenbrown}{rgb}{0.6, 0.4, 0.08}
\definecolor{davysgrey}{rgb}{0.33, 0.33, 0.33}
\definecolor{deepsaffron}{rgb}{1.0, 0.6, 0.2}

\begin{document}

\begin{frontmatter}

\title{On the use of case estimate and transactional payment data in neural networks for individual loss reserving}

\cortext[cor]{Corresponding author.}

\address[UMelb]{Centre for Actuarial Studies, Department of Economics, University of Melbourne VIC 3010, Australia}
\address[UNSW]{School of Risk and Actuarial Studies, UNSW Business School, UNSW Sydney NSW 2052, Australia}

\author[UMelb]{Benjamin Avanzi}
\ead{b.avanzi@unimelb.edu.au}

\author[UMelb]{Matthew Lambrianidis\corref{cor}}
\ead{matthew.lambrianidis@gmail.com}

\author[UNSW]{Greg Taylor}
\ead{greg\_taylor60@hotmail.com}

\author[UNSW]{Bernard Wong}
\ead{bernard.wong@unsw.edu.au}

\begin{abstract}

The use of neural networks trained on individual claims data has become increasingly popular in the actuarial reserving literature. We consider how to best input historical payment data in neural network models. Additionally, case estimates are also available in the format of a time series, and we extend our analysis to assessing their predictive power.

In this paper, we compare a feed-forward neural network trained on summarised transactions to a recurrent neural network equipped to analyse a claim’s entire payment history and/or case estimate development history. We draw conclusions from training and comparing the performance of the models on multiple, comparable highly complex datasets simulated from SPLICE \citep*{AvTaWa23}. In environments where covariate data is limited, we find evidence that case estimate data will improve predictions substantially, but that equipping the neural network with memory only leads to meagre improvements. Although the case estimation process and quality will vary significantly between insurers, we provide a standardised methodology for assessing their value.

\end{abstract}

\begin{keyword} Loss Reserving, Neural Networks, Case Estimates, Individual Claims, Deep Learning, RBNS Reserves

JEL codes:  
G22	\sep 
C45	\sep 
C53	

MSC classes:
91G70 \sep 	
91G60 \sep 	
62P05 \sep 	
91B30 

\end{keyword}
\end{frontmatter}

\section{Introduction}

\subsection{Background and motivations}

Forecasting reserves for the outstanding claims liabilities is one of the main challenges consistently faced by insurers. Although it is a regulatory requirement - see for instance GPS 340\footnote{\url{https://www.legislation.gov.au/F2023L00699/latest/text}} in Australia and Solvency II\footnote{\url{https://www.eiopa.europa.eu/browse/regulation-and-policy/solvency-ii_en}} in Europe - associated benefits also extend to pricing and general capital management \citep{Tay19}. Therefore, it is in the interest of all insurers to invest significant resources into their claims models.

Historically, models that use aggregated claims data (so-called ``triangles'') have been more popular than those that use individual claims data due to their relative ease of calculation. Recently, however, this trend has been reversing. With ever-increasing computing capabilities, it is becoming increasingly feasible to produce estimates at an individual claim level in a timely manner. The key advantage of doing so is to gain access to data that is lost upon aggregation \citep*{TaMcSu08}. At a minimum, this would include information about an individual claim's payment history, but it may also include features such as the age, gender or occupation of the policyholder, any written claims descriptions, and any information about the case estimates and potential revisions for each claim. Having access to these features may lead to more robust estimates of individual claims, which in turn could lead to more accurate estimates for the group. \citet*{HuWuZh16} demonstrate that in portfolios where only a single payment is made per claim, individual methods asymptotically outperform aggregate methods such as the chain ladder and Bornhuetter-Ferguson. Although the data used in our paper contains multiple payments per claim, having this result provides an incentive to explore individual models more deeply. 

In addition to the shift from aggregate to individual models, there has also been a gradual movement towards machine learning models. These range from GLMs \citep*{TaMcSu08, TaMc16, AvLiWoXi24} to tree-based methods \citep*{LoMiTh19, DuPi19, Wut18b}, reinforcement learning \citep*{AvRiWoWuXi25}, and neural network variants \citep*{Mul06, Gab21, AMAvTaWo22}. The primary benefit of these more advanced models is their ability to model relatively complex relationships that exist in the realm of claims modelling, with tree-based methods and neural networks having the greatest flexibility. These models tend to be most useful in environments that violate the assumptions made by simpler models, such as GLMs and the chain ladder. The trade-off, however, is a lack of interpretability in the forecasts produced by the models \citep{Bre01b}.

There are cases where both interpretability (and the related concept of explainability; see \cite*{LaPhWo25}) and complexity are necessary. One of the primary reasons is due to regulatory requirements. This has led to the emergence of `boosted' models that begin with a relatively explainable model such as a GLM, and make fine adjustments through the use of a tree-based or neural network model. See \citet*{ScWu19} and \citet*{GaRiWu19} for examples using a neural network. In this paper, however, interpretability is not considered.

Our goal is instead to investigate the relative merits of two common variants of neural networks in an individual loss reserving context, particularly where transaction-level payment and case estimate data are available and where other covariate data is limited. Our interest is whether the recurrent structure (with ``memory'') adds value over a more straightforward `feed-forward' structure, rather than whether these models are ``state-of-the-art''. To illustrate the impact of including these data, we focus on the simplest neural network structures that allow for them, notably a feed-forward and a recurrent network.

\subsection{Contributions}

In line with the developments in neural network architectures and their applications with actuarial work, the literature applying Recurrent Neural Networks (RNNs) in a reserving context has been gaining traction in recent years: \citet*{Kuo19} uses them in the construction of the `DeepTriangle' model for aggregate predictions, with \citet*{CaAbJe25} extending this model to account for the dependence between different lines of business. \citet*{ChBeCoCo23} instead utilise an RNN variant to forecast claim amounts at an individual level, the general approach later refined by \citet*{ScSc25}. However, to date none of the existing literature quantifies the additional value of an RNN relative to a feed-forward neural network - or if there is any benefit - in a loss reserving context. And indeed, there has been very limited discussion on the theoretical motivations for using an RNN, particularly over comparable models such as a feed-forward network, in this context. Given the increased traction that both of these models are gaining in the reserving space, it seems particularly important to address this significant gap. This will form the basis of our first research question.

As previously mentioned, individual claims models have been extensively covered in the literature in recent years. However, case estimate data is largely absent from these papers. \citet*{TaMcSu08} incorporate individual case estimate data in a GLM setting, whereas \citet*{PiAnDe14} and \citet*{MeWu10} do so using a `Paid-Incurred Chain' model: \citeauthor*{MeWu10} with aggregate data, and \citeauthor*{PiAnDe14} with individual data. Notably, none of these papers use machine learning models. Hence, one of our main contributions lies in the combination of individual case estimates with machine learning techniques.

The inclusion of case estimate data will allow us to explore its relative merits by comparing the results of the model that is trained with case estimate data to the model trained without it. Since the case estimation process varies significantly between insurers, we do not intend to answer to this question definitively in the paper. Rather, we want to provide a framework that can be used by insurers to assess the value of their own case estimates. Where covariate data is plentifully available, the case estimate data may implicitly encode some of the patterns that can be extracted directly from the covariates. Case estimate data is therefore likely to be of higher importance in environments where these covariates are sparse, such as the environment examined throughout this paper. Consequently, any empirical findings presented here will be more relevant for insurers with available case estimate data but limited covariate data.

The key focus areas of this paper can then be summarised into the following three research questions:

\begin{enumerate}
    \item Is there a significant difference in performance between a feed-forward network trained solely on summary data compared to an RNN trained on full transactional data?
    \item Does a neural network (recurrent or feed-forward) trained with case estimate data perform better than the same model without case estimate data?
    \item Does a neural network (recurrent or feed-forward) trained without case estimate data outperform the case estimates on their own?
\end{enumerate} To our knowledge, there are no other papers in the reserving literature that answer questions 1 or 3. Question 2 is considered by both \citet*{TaMcSu08} and \citet*{PiAnDe14}, however neither do so in a neural network context. Therefore, a gap exists in the literature as, at the time of writing, no other existing work has compared the performance of an RNN to a feed-forward network, or thoroughly examined the usefulness of case estimate data, in the context of individual loss reserving with neural networks.

We will address these questions by designing an environment that fairly assesses each model, while also ensuring the validity of the case estimates as a ``model''. Particular focus is placed on the use of practical and realistic data processing, including, but not limited to, the splitting of datasets between training, validation and test sets. We use both quantitative and qualitative results that assess model performance at an aggregate and an individual claim level to answer each question. Model performance is evaluated across multiple datasets generated by the SPLICE simulator \citep*{AvTaWaWo21,AvTaWa23}. While these answers will not necessarily be definitive, the general framework can be used to conduct a similar assessment in an environment that is applicable to a given insurer.

\subsection{Outline of Paper}

Section \ref{Model Architectures} defines the architectures of our neural network models, while Section \ref{Training Methodology} explores the training process and associated methodology. The results are contained in Section \ref{Results}. Section \ref{Conclusion} concludes. Additional results and technical details are located in the Appendix.

\section{Model architectures} \label{Model Architectures}

Like all supervised machine learning models, a neural network takes an input $X$, applies a function $f$ and returns an output $y$ - the choice of $y$ being one of the key challenges in individual loss reserving. This paper will consider two different forms of neural network architectures. The primary reason is for ease of comparison. However, we also acknowledge the theoretical benefits of the `Universal Approximation Theorem' \citep{Cyb89}, which proves that it is possible for a feed-forward neural network with a single hidden layer to approximate any continuous and compactly supported function. Although this is not guaranteed in a practical setting, neural networks do still tend to be favoured in highly complex environments where other machine learning models struggle.

In this section, we begin with an overview and comparison of our models, then discuss some of the key modelling choices.

\begin{figure}[!htb]
    \centering
    \scalebox{.9}{
    \begin{tikzpicture}[
    roundnode/.style={ellipse, draw=black, thick, fill=green!10, minimum width=17mm, minimum height=2mm, align=center},
    circlenode/.style={circle, draw=black, thick, fill=blue!10, minimum size=10mm, align=center},
    group/.style={draw, dashed, thick, fill=none, inner sep=5pt}
    ]
    
    \node[roundnode, fill=deepsaffron!50] (cal quarter) {\footnotesize cal quarter};
    \node[roundnode, fill=deepsaffron!50] (dev quarter) [below=2mm of cal quarter] {\footnotesize dev quarter};
    \node[roundnode] (acc quarter) [below=2mm of dev quarter] {\footnotesize acc quarter};
    \node[roundnode, fill=deepsaffron!50] (payment summaries) [below=2mm of acc quarter] {\footnotesize payment summaries};
    \node[roundnode, fill=yellow!30] (case estimate summaries) [below=2mm of payment summaries] {\footnotesize case estimate summaries};
    \node[roundnode] (legal rep) [below=2mm of case estimate summaries] {\footnotesize legal rep};
    \node[roundnode, fill=violet!10] (embedding1) [below=2mm of legal rep] {\footnotesize embedding};
    \node[roundnode, fill=violet!10] (embedding2) [below=2mm of embedding1] {\footnotesize embedding};
    \node[roundnode] (severity) [left=5mm of embedding1] {\footnotesize severity};
    \node[roundnode] (age) [left=5mm of embedding2] {\footnotesize age};

    \node[circlenode] (fnn11) [right = 35mm of dev quarter] {};
    \node[circlenode] (fnn12) [below = 5mm of fnn11] {};
    \begin{scope}[shift={(fnn12.south)}]
        \foreach \y in {-0.8, -1, -1.2}  
            \fill (0, \y) circle (1.5pt);
        \coordinate (fnn1vdot) at (0.5, -1);
    \end{scope}
    \node[circlenode] (fnn13) [below = 20mm of fnn12] {};

    \begin{scope}[shift={(fnn11.east)}] 
        \foreach \x in {0.8, 1, 1.2}  
            \fill (\x,0) circle (1.5pt);
        \coordinate (fnnhdot1 west) at (0.7, 0);
        \coordinate (fnnhdot1 east) at (1.3, 0);
    \end{scope}
    \begin{scope}[shift={(fnn12.east)}] 
        \foreach \x in {0.8, 1, 1.2} 
            \fill (\x,0) circle (1.5pt);
        \coordinate (fnnhdot2 west) at (0.7, 0);
        \coordinate (fnnhdot2 east) at (1.3, 0);
    \end{scope}
    \begin{scope}[shift={(fnn1vdot.east)}]
        \foreach \x in {0.8, 1, 1.2}  
            \fill (\x,0) circle (1.5pt);
    \end{scope}
    \begin{scope}[shift={(fnn13.east)}]
        \foreach \x in {0.8, 1, 1.2} 
            \fill (\x,0) circle (1.5pt);
        \coordinate (fnnhdot3 west) at (0.7, 0);
        \coordinate (fnnhdot3 east) at (1.3, 0);
    \end{scope}

    \node[circlenode] (fnn21) [right = 20mm of fnn11] {};
    \node[circlenode] (fnn22) [below = 5mm of fnn21] {};
    \begin{scope}[shift={(fnn22.south)}]
        \foreach \y in {-0.8, -1, -1.2} 
            \fill (0, \y) circle (1.5pt);
        \coordinate (fnn2vdot) at (0.6, -1);
    \end{scope}
    \node[circlenode] (fnn23) [below = 20mm of fnn22] {};

    \node[circlenode, fill=davysgrey!10] (output) [right=80mm of case estimate summaries] {};

    \node[group, fit=(fnn11) (fnn13) (fnn23), label={[text=blue] above:Feed-forward layers}, color=blue] (G2) {};
    \node[group, fit=(output), label={[text=davysgrey] above:Output}, color=davysgrey] (G3) {};
    \node[group, fit=(cal quarter) (age) (payment summaries), label={[text=goldenbrown] above:Static inputs}, color=goldenbrown] (G3) {};

    \draw[->] (severity) -- (embedding1);
    \draw[->] (age) -- (embedding2);

    \draw[->] (cal quarter) -- (fnn11);
    \draw[->] (dev quarter) -- (fnn11);
    \draw[->] (acc quarter) -- (fnn11);
    \draw[->] (payment summaries) -- (fnn11);
    \draw[->] (case estimate summaries) -- (fnn11);
    \draw[->] (legal rep) -- (fnn11);
    \draw[->] (embedding1) -- (fnn11);
    \draw[->] (embedding2) -- (fnn11);

    \draw[->] (cal quarter) -- (fnn12);
    \draw[->] (dev quarter) -- (fnn12);
    \draw[->] (acc quarter) -- (fnn12);
    \draw[->] (payment summaries) -- (fnn12);
    \draw[->] (case estimate summaries) -- (fnn12);
    \draw[->] (legal rep) -- (fnn12);
    \draw[->] (embedding1) -- (fnn12);
    \draw[->] (embedding2) -- (fnn12);

    \draw[->] (cal quarter) -- (fnn13);
    \draw[->] (dev quarter) -- (fnn13);
    \draw[->] (acc quarter) -- (fnn13);
    \draw[->] (payment summaries) -- (fnn13);
    \draw[->] (case estimate summaries) -- (fnn13);
    \draw[->] (legal rep) -- (fnn13);
    \draw[->] (embedding1) -- (fnn13);
    \draw[->] (embedding2) -- (fnn13);

    \draw[->] (fnn11) -- (fnnhdot1 west);
    \draw[->] (fnn11) -- (fnnhdot2 west);
    \draw[->] (fnn11) -- (fnnhdot3 west);
    
    \draw[->] (fnn12) -- (fnnhdot1 west);
    \draw[->] (fnn12) -- (fnnhdot2 west);
    \draw[->] (fnn12) -- (fnnhdot3 west);
    
    \draw[->] (fnn13) -- (fnnhdot1 west);
    \draw[->] (fnn13) -- (fnnhdot2 west);
    \draw[->] (fnn13) -- (fnnhdot3 west);

    \draw[->] (fnnhdot1 east) -- (fnn21);
    \draw[->] (fnnhdot1 east) -- (fnn22);
    \draw[->] (fnnhdot1 east) -- (fnn23);

    \draw[->] (fnnhdot2 east) -- (fnn21);
    \draw[->] (fnnhdot2 east) -- (fnn22);
    \draw[->] (fnnhdot2 east) -- (fnn23);

    \draw[->] (fnnhdot3 east) -- (fnn21);
    \draw[->] (fnnhdot3 east) -- (fnn22);
    \draw[->] (fnnhdot3 east) -- (fnn23);

    \draw[->] (fnn21) -- (output);
    \draw[->] (fnn22) -- (output);
    \draw[->] (fnn23) -- (output);
        
    \end{tikzpicture}
    } 

    \caption{Architecture of FNN and FNN+ models (as will be defined in Section \ref{sec:inputs}). The yellow node (case estimate summaries) is included in the FNN+ and excluded from the FNN. All inputs are static.}
    \label{fig:FNN Architecture}

    \scalebox{.9}{
    \begin{tikzpicture}[
    roundnode/.style={ellipse, draw=black, thick, fill=green!10, minimum width=17mm, minimum height=2mm, align=center},
    squarednode/.style={rectangle, draw=black, thick, minimum size=20mm, align=center},
    circlenode/.style={circle, draw=black, thick, fill=blue!10, minimum size=10mm, align=center},
    group/.style={draw, dashed, thick, fill=none, inner sep=5pt}
    ]
    
    \node[roundnode, fill=deepsaffron!50] (cal)                     {\footnotesize cal time};
    \node[roundnode, fill=deepsaffron!50] (dev) [below=2mm of cal]   {\footnotesize dev time};
    \node[roundnode, fill=deepsaffron!50] (paid) [below=2mm of dev] {\footnotesize cumpaid};
    \node[roundnode, fill=yellow!30] (case estimate) [below=2mm of paid] {\footnotesize case estimate};

    \node[circlenode, fill=magenta!20] (rnn11) [above right = 10mm and 15mm of cal] {};
    \draw[->, thick, looseness=4, out=40, in=80] (rnn11) to (rnn11);
    
    \node[circlenode, fill=magenta!20] (rnn12) [below = 4mm of rnn11] {};
    \draw[->, thick, looseness=4, out=40, in=80] (rnn12) to (rnn12);
    
    \begin{scope}[shift={(rnn12.south)}]
        \foreach \y in {-0.4, -0.6, -0.8} 
            \fill (0, \y) circle (1.5pt);
        \coordinate (rnn1vdot) at (0.5, -0.8);
    \end{scope}
    \node[circlenode, fill=magenta!20] (rnn13) [below = 12mm of rnn12] {};
    \draw[->, thick, looseness=4, out=40, in=80] (rnn13) to (rnn13);

    \begin{scope}[shift={(rnn11.east)}]
        \foreach \x in {0.8, 1, 1.2} 
            \fill (\x,0) circle (1.5pt);
        \coordinate (rnnhdot1 west) at (0.7, 0);
        \coordinate (rnnhdot1 centre) at (1, 0);
        \coordinate (rnnhdot1 east) at (1.3, 0);
    \end{scope}
    \begin{scope}[shift={(rnn12.east)}] 
        \foreach \x in {0.8, 1, 1.2}  
            \fill (\x,0) circle (1.5pt);
        \coordinate (rnnhdot2 west) at (0.7, 0);
        \coordinate (rnnhdot2 east) at (1.3, 0);
    \end{scope}
    \begin{scope}[shift={(rnn1vdot.east)}] 
        \foreach \x in {0.8, 1, 1.2}  
            \fill (\x,0) circle (1.5pt);
    \end{scope}
    \begin{scope}[shift={(rnn13.east)}] 
        \foreach \x in {0.8, 1, 1.2} 
            \fill (\x,0) circle (1.5pt);
        \coordinate (rnnhdot3 west) at (0.7, 0);
        \coordinate (rnnhdot3 east) at (1.3, 0);
    \end{scope}

    \node[circlenode, fill=magenta!20] (rnn21) [right = 20mm of rnn11] {};
    \draw[->, thick, looseness=4, out=40, in=80] (rnn21) to (rnn21);
    
    \node[circlenode, fill=magenta!20] (rnn22) [below = 4mm of rnn21] {};
    \draw[->, thick, looseness=4, out=40, in=80] (rnn22) to (rnn22);
    
    \begin{scope}[shift={(rnn22.south)}] 
        \foreach \y in {-0.4, -0.6, -0.8} 
            \fill (0, \y) circle (1.5pt);
        \coordinate (rnn2vdot) at (0.6, -1);
    \end{scope}
    
    \node[circlenode, fill=magenta!20] (rnn23) [below = 12mm of rnn22] {};
    \draw[->, thick, looseness=4, out=40, in=80] (rnn23) to (rnn23);

    \node[roundnode] (cal quarter) [below=4mm of rnn23] {\footnotesize cal quarter};
    \node[roundnode] (acc quarter) [below=2mm of cal quarter] {\footnotesize acc quarter};
    \node[roundnode] (legal rep) [below=2mm of acc quarter] {\footnotesize legal rep};
    \node[roundnode, fill=violet!10] (embedding1) [below=2mm of legal rep] {\footnotesize embedding};
    \node[roundnode, fill=violet!10] (embedding2) [below=2mm of embedding1] {\footnotesize embedding};
    \node[roundnode] (severity) [left=10mm of embedding1] {\footnotesize severity};
    \node[roundnode] (age) [left=10mm of embedding2] {\footnotesize age};

    \node[circlenode] (fnn11) [right = 25mm of rnn2vdot] {};
    \node[circlenode] (fnn12) [below = 5mm of fnn11] {};
    \begin{scope}[shift={(fnn12.south)}] 
        \foreach \y in {-0.8, -1, -1.2} 
            \fill (0, \y) circle (1.5pt);
        \coordinate (fnn1vdot) at (0.5, -1);
    \end{scope}
    \node[circlenode] (fnn13) [below = 20mm of fnn12] {};

    \begin{scope}[shift={(fnn11.east)}] 
        \foreach \x in {0.8, 1, 1.2} 
            \fill (\x,0) circle (1.5pt);
        \coordinate (fnnhdot1 west) at (0.7, 0);
        \coordinate (fnnhdot1 centre) at (1, 0);
        \coordinate (fnnhdot1 east) at (1.3, 0);
    \end{scope}
    \begin{scope}[shift={(fnn12.east)}]
        \foreach \x in {0.8, 1, 1.2}  
            \fill (\x,0) circle (1.5pt);
        \coordinate (fnnhdot2 west) at (0.7, 0);
        \coordinate (fnnhdot2 east) at (1.3, 0);
    \end{scope}
    \begin{scope}[shift={(fnn1vdot.east)}] 
        \foreach \x in {0.8, 1, 1.2}  
            \fill (\x,0) circle (1.5pt);
    \end{scope}
    \begin{scope}[shift={(fnn13.east)}]
        \foreach \x in {0.8, 1, 1.2}  
            \fill (\x,0) circle (1.5pt);
        \coordinate (fnnhdot3 west) at (0.7, 0);
        \coordinate (fnnhdot3 east) at (1.3, 0);
    \end{scope}

    \node[circlenode] (fnn21) [right = 20mm of fnn11] {};
    \node[circlenode] (fnn22) [below = 5mm of fnn21] {};
    \begin{scope}[shift={(fnn22.south)}]
        \foreach \y in {-0.8, -1, -1.2}   
            \fill (0, \y) circle (1.5pt);
        \coordinate (fnn2vdot) at (0.6, -1);
    \end{scope}
    \node[circlenode] (fnn23) [below = 20mm of fnn22] {};

    \node[circlenode, fill=davysgrey!10] (output) [right=70mm of acc quarter] {};

    \node[group, fit=(rnn11) (rnn13) (rnn23), label={[text=magenta] above:Recurrent layers}, color=magenta] (G1) {};
    \node[group, fit=(fnn11) (fnn13) (fnn23), label={[text=blue] above:Feed-forward layers}, color=blue] (G2) {};
    \node[group, fit=(output), label={[text=davysgrey] above:Output}, color=davysgrey] (G3) {};
    \node[group, fit=(cal) (case estimate), label={[text=darkspringgreen] above:Time series inputs}, color=darkspringgreen] (G3) {};
    \node[group, fit=(cal quarter) (age), label={[text=goldenbrown] below:Static inputs}, color=goldenbrown] (G3) {};

    \draw[->] (severity) -- (embedding1);
    \draw[->] (age) -- (embedding2);
    
    \draw[->] (cal) -- (rnn11);
    \draw[->] (dev) -- (rnn11);
    \draw[->] (paid) -- (rnn11);
    \draw[->] (case estimate) -- (rnn11);

    \draw[->] (cal) -- (rnn12);
    \draw[->] (dev) -- (rnn12);
    \draw[->] (paid) -- (rnn12);
    \draw[->] (case estimate) -- (rnn12);

    \draw[->] (cal) -- (rnn13);
    \draw[->] (dev) -- (rnn13);
    \draw[->] (paid) -- (rnn13);
    \draw[->] (case estimate) -- (rnn13);

    \draw[->] (rnn11) -- (rnnhdot1 west);
    \draw[->] (rnn11) -- (rnnhdot2 west);
    \draw[->] (rnn11) -- (rnnhdot3 west);
    
    \draw[->] (rnn12) -- (rnnhdot1 west);
    \draw[->] (rnn12) -- (rnnhdot2 west);
    \draw[->] (rnn12) -- (rnnhdot3 west);
    
    \draw[->] (rnn13) -- (rnnhdot1 west);
    \draw[->] (rnn13) -- (rnnhdot2 west);
    \draw[->] (rnn13) -- (rnnhdot3 west);

    \draw[->] (rnnhdot1 east) -- (rnn21);
    \draw[->] (rnnhdot1 east) -- (rnn22);
    \draw[->] (rnnhdot1 east) -- (rnn23);

    \draw[->] (rnnhdot2 east) -- (rnn21);
    \draw[->] (rnnhdot2 east) -- (rnn22);
    \draw[->] (rnnhdot2 east) -- (rnn23);

    \draw[->] (rnnhdot3 east) -- (rnn21);
    \draw[->] (rnnhdot3 east) -- (rnn22);
    \draw[->] (rnnhdot3 east) -- (rnn23);

    \draw[->] (rnn21) -- (fnn11);
    \draw[->] (rnn21) -- (fnn12);
    \draw[->] (rnn21) -- (fnn13);
    \draw[->] (rnn22) -- (fnn11);
    \draw[->] (rnn22) -- (fnn12);
    \draw[->] (rnn22) -- (fnn13);
    \draw[->] (rnn23) -- (fnn11);
    \draw[->] (rnn23) -- (fnn12);
    \draw[->] (rnn23) -- (fnn13);

    \draw[->] (cal quarter) -- (fnn11);
    \draw[->] (cal quarter) -- (fnn12);
    \draw[->] (cal quarter) -- (fnn13);
    \draw[->] (acc quarter) -- (fnn11);
    \draw[->] (acc quarter) -- (fnn12);
    \draw[->] (acc quarter) -- (fnn13);
    \draw[->] (legal rep) -- (fnn11);
    \draw[->] (legal rep) -- (fnn12);
    \draw[->] (legal rep) -- (fnn13);
    \draw[->] (embedding1) -- (fnn11);
    \draw[->] (embedding1) -- (fnn12);
    \draw[->] (embedding1) -- (fnn13);
    \draw[->] (embedding2) -- (fnn11);
    \draw[->] (embedding2) -- (fnn12);
    \draw[->] (embedding2) -- (fnn13);
    
    \draw[->] (fnn11) -- (fnnhdot1 west);
    \draw[->] (fnn11) -- (fnnhdot2 west);
    \draw[->] (fnn11) -- (fnnhdot3 west);
    
    \draw[->] (fnn12) -- (fnnhdot1 west);
    \draw[->] (fnn12) -- (fnnhdot2 west);
    \draw[->] (fnn12) -- (fnnhdot3 west);
    
    \draw[->] (fnn13) -- (fnnhdot1 west);
    \draw[->] (fnn13) -- (fnnhdot2 west);
    \draw[->] (fnn13) -- (fnnhdot3 west);

    \draw[->] (fnnhdot1 east) -- (fnn21);
    \draw[->] (fnnhdot1 east) -- (fnn22);
    \draw[->] (fnnhdot1 east) -- (fnn23);

    \draw[->] (fnnhdot2 east) -- (fnn21);
    \draw[->] (fnnhdot2 east) -- (fnn22);
    \draw[->] (fnnhdot2 east) -- (fnn23);

    \draw[->] (fnnhdot3 east) -- (fnn21);
    \draw[->] (fnnhdot3 east) -- (fnn22);
    \draw[->] (fnnhdot3 east) -- (fnn23);

    \draw[->] (fnn21) -- (output);
    \draw[->] (fnn22) -- (output);
    \draw[->] (fnn23) -- (output);

    \end{tikzpicture}
    } 

    \caption{Architecture of LSTM and LSTM+ models (as will be defined in Section \ref{sec:inputs}). The yellow node (case estimate) is included in the LSTM+ and excluded from the LSTM. The inputs to the recurrent layer(s) are time series, while the remaining inputs are static.}
    \label{fig:LSTM Architecture}
\end{figure}

\subsection{Feed-Forward Networks} \label{sec:feed-forward}

Feed-forward networks are a fundamental and widely-used variant of neural networks \citep{WuRiAvLiMaScJuSc25, EmWu22}. They are built from a collection of units (also known as nodes), which are non-linear transformations of linear combinations of the inputs. A feed-forward network usually contains multiple units per layer, with each unit learning its own weights and bias terms. As Figures \ref{fig:FNN Architecture} and \ref{fig:LSTM Architecture} suggest, the units in layer $l$ become the inputs for the units in layer $l + 1$.

To illustrate these concepts, we define the following notation:

\begin{itemize}
    \item Let $L_F$ be the number of feed-forward layers. $p^{(l)}$ is the output dimension for layer $l \in \{1, 2, ..., L_F\}$, which is also the input dimension for layer $l+1$. $p^{(0)}$ represents the number of features in the input layer.
    \item $\mathbf{x}_j = (x_{j,1}, x_{j,2}, ..., x_{j,p^{(0)}})^T \in \mathbb{R}^{p^{(0)}} $ are the inputs to the first feed-forward layer for observation $j$.
    \item $y_j \in \mathbb{R}$ is the final output of the network for observation $j$.
    \item $\phi^l(\cdot)$ is the activation function in the $l$\textsuperscript{th} layer.
\end{itemize} Set $\mathbf{z}^{0}_j = \mathbf{x}_j$. Then the following equation represents the mapping from layer $l$ to $l+1$ for observation $j$: 

\begin{equation} \label{E_FNN}
    \mathbf{z}^{l+1}_j = \phi^{l+1}\left(\mathbf{W}^{l+1}\mathbf{z}^l_j + \mathbf{b}^{l+1}\right), \quad l \in \{0, 1, ..., L_F-1 \},
\end{equation} where $\mathbf{b}^l \in \mathbb{R}^{p^{(l)}}$ are bias terms for each unit in layer $l$ and $\mathbf{W}^l \in \mathbb{R}^{p^{(l)} \times p^{(l-1)}}$ are the weights for each (output, input) combination in layer $l$.

\subsection{Recurrent Neural Networks} \label{sec:rnn}

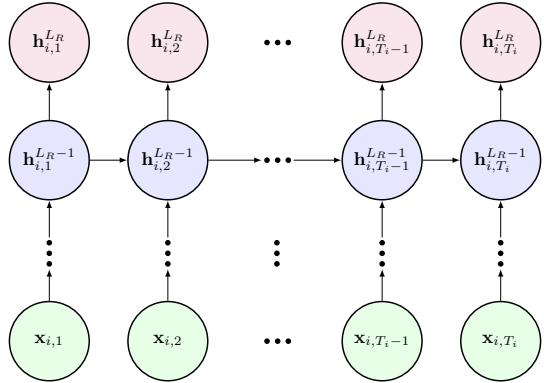
\begin{figure}[!htb]
    \centering
    \scalebox{.7}{
    \begin{tikzpicture}[
    roundnode/.style={circle, draw=black, thick, fill=green!10, minimum size=15mm, align=center},
    squarednode/.style={rectangle, draw=black, thick, minimum size=20mm, align=center},
    ]
    
    \node[roundnode] (x1)                     {$\mathbf{x}_{j,1}$};
    \node[roundnode] (x2) [right=7mm of x1]   {$\mathbf{x}_{j,2}$};
    \node[roundnode] (xn-1) [right=25mm of x2] {$\mathbf{x}_{j,T_i-1}$};
    \node[roundnode] (xn) [right=7mm of xn-1] {$\mathbf{x}_{j,T_i}$};

    \begin{scope}[shift={(x1.north)}]
        \foreach \y in {0.7, 0.9, 1.1} 
            \fill (0, \y) circle (1.5pt);
        \coordinate (x1vdot north) at (0, 1.2);
        \coordinate (x1vdot south) at (0, 0.6);
    \end{scope}

    \begin{scope}[shift={(x2.north)}]
        \foreach \y in {0.7, 0.9, 1.1} 
            \fill (0, \y) circle (1.5pt);
        \coordinate (x2vdot north) at (0, 1.2);
        \coordinate (x2vdot south) at (0, 0.6);
        \coordinate (x2vdot east) at (2.05, 0);
    \end{scope}

    \begin{scope}[shift={(x2vdot east.east)}]
        \foreach \y in {0.7, 0.9, 1.1} 
            \fill (0, \y) circle (1.5pt);
    \end{scope}

    \begin{scope}[shift={(xn-1.north)}]
        \foreach \y in {0.7, 0.9, 1.1} 
            \fill (0, \y) circle (1.5pt);
        \coordinate (xn-1vdot north) at (0, 1.2);
        \coordinate (xn-1vdot south) at (0, 0.6);
    \end{scope}

    \begin{scope}[shift={(xn.north)}]
        \foreach \y in {0.7, 0.9, 1.1} 
            \fill (0, \y) circle (1.5pt);
        \coordinate (xnvdot north) at (0, 1.2);
        \coordinate (xnvdot south) at (0, 0.6);
    \end{scope}

    \node[roundnode, fill=blue!10] (r1) [above=7mm of x1vdot north]   {$\mathbf{h}^{L_R-1}_{j,1}$};
    \node[roundnode, fill=blue!10] (r2) [right=7mm of r1]   {$\mathbf{h}^{L_R-1}_{j,2}$};
    \node[roundnode, fill=blue!10] (rn-1) [right=25mm of r2] {$\mathbf{h}^{L_R-1}_{j,T_i-1}$};
    \node[roundnode, fill=blue!10] (rn) [right=7mm of rn-1] {$\mathbf{h}^{L_R-1}_{j,T_i}$};

    \node[roundnode, fill=purple!10] (h1) [above=7mm of r1]   {$\mathbf{h}^{L_R}_{j,1}$};
    \node[roundnode, fill=purple!10] (h2) [right=7mm of h1]   {$\mathbf{h}^{L_R}_{j,2}$};
    \node[roundnode, fill=purple!10] (hn-1) [right=25mm of h2] {$\mathbf{h}^{L_R}_{j,T_i-1}$};
    \node[roundnode, fill=purple!10] (hn) [right=7mm of hn-1] {$\mathbf{h}^{L_R}_{j,T_i}$};

    \begin{scope}[shift={(x2.east)}]
        \foreach \x in {1.1, 1.3, 1.5}  
            \fill (\x,0) circle (1.5pt);
        \coordinate (xdot west) at (1, 0);
        \coordinate (xdot east) at (1.6, 0);
    \end{scope}

    \begin{scope}[shift={(r2.east)}]
        \foreach \x in {1.1, 1.3, 1.5}  
            \fill (\x,0) circle (1.5pt);
        \coordinate (rdot west) at (1, 0);
        \coordinate (rdot east) at (1.6, 0);
    \end{scope}

    \begin{scope}[shift={(h2.east)}]
        \foreach \x in {1.1, 1.3, 1.5}  
            \fill (\x,0) circle (1.5pt);
        \coordinate (hdot west) at (1, 0);
        \coordinate (hdot east) at (1.6, 0);
    \end{scope}
    
    \draw[->] (x1) -- (x1vdot south);
    \draw[->] (x2) -- (x2vdot south);
    \draw[->] (xn-1) -- (xn-1vdot south);
    \draw[->] (xn) -- (xnvdot south);

    \draw[->] (x1vdot north) -- (r1);
    \draw[->] (x2vdot north) -- (r2);
    \draw[->] (xn-1vdot north) -- (rn-1);
    \draw[->] (xnvdot north) -- (rn);

    \draw[->] (r1) -- (h1);
    \draw[->] (r2) -- (h2);
    \draw[->] (rn-1) -- (hn-1);
    \draw[->] (rn) -- (hn);

    \draw[->] (r1) -- (r2);
    \draw[->] (r2) -- (rdot west);
    \draw[->] (rdot east) -- (rn-1);
    \draw[->] (rn-1) -- (rn);
        
    \end{tikzpicture}
    } 

    \caption{Example of an unrolled RNN.}
    \label{fig:RNN}
\end{figure}

\noindent An alternative to feed-forward networks, RNNs are a particular class of neural network that are distinguished by their recursive structure, as seen in Figure \ref{fig:LSTM Architecture}. Similarly to Section \ref{sec:feed-forward}, we define the following notation for a basic recurrent layer:
\begin{itemize}
    \item Let $L_R$ be the number of recurrent layers. $p^{(l)}$ is the output dimension for layer $l \in \{1, 2, ..., L_R\}$, which is also the input dimension for layer $l+1$. $p^{(0)}$ represents the number of concurrent time series in the input layer.
    \item $T_j$ is the sequence length for observation $j$, noting that the sequence length can differ between observations. $t \in \{1, 2, ..., T_j\}$ is the value of the current time step.
    \item $\mathbf{x}_{j,t} = (x_{j,1,t}, x_{j,2,t}, ..., x_{j,p^{(0)},t})^T \in \mathbb{R}^{p^{(0)}}$ are the inputs to the first recurrent layer for observation $j$.
    \item $\phi^l(\cdot)$ is the activation function in the $l$\textsuperscript{th} layer.
\end{itemize} 
Set $\mathbf{h}^{0}_{j,t} = \mathbf{x}_{j,t}$. Then, the recurrent layers can be represented by the following equation:
\begin{equation} \label{E_RNN}
    \mathbf{h}^{l+1}_{j,t} = \phi^{l+1}\left(\mathbf{W}^{l+1}_{hh} \mathbf{h}^{l+1}_{j,t-1} + \mathbf{W}^{l+1}_{xh} \mathbf{h}^l_{j,t} + \mathbf{b}^{l+1}\right), \quad l \in \{0, 1, ..., L_R-1 \}
\end{equation} where $\mathbf{W}^l_{hh} \in \mathbb{R}^{p^{(l)} \times p^{(l)}}$ and $\mathbf{W}^l_{xh} \in \mathbb{R}^{p^{(l)} \times p^{(l-1)}}$ both represent sets of weights, and $\mathbf{b}^l \in \mathbb{R}^{p^{(l)}}$ are the bias terms for each unit in layer $l$. Figure \ref{fig:RNN} demonstrates the operation of an RNN in an `unrolled' graphical form, illustrating how information from past transactions propagates forward through time.

Although an output is produced at each time step, the first feed-forward layer in Figure \ref{fig:LSTM Architecture} only relies on the output at the final time step from the final recurrent layer. This is because it is the only output that depends on the inputs at every time step. However, for relatively long sequences, the RNN can struggle to `remember' information from the beginning of the sequence. This issue is known as the `Vanishing Gradient Problem' \citep*{RuHiWi86}.

The Long-Short Term Memory (LSTM) network \citep*{HoSc97} is one such RNN variation that deals with the vanishing gradient problem. It includes a memory cell, which stores long-term information through the use of three gates: input, forget, and output gates. At each time step $t$, the input gate determines what proportion of the input at $t$ should be added to the memory cell. The forget gate determines the proportion of the memory cell to discard. Finally, the output gate determines how much of the memory cell should be passed on to the next hidden state. Through these gates, the LSTM is able to store important information in the memory cell to access at a later time step, while a `vanilla' RNN will likely forget that information before that time step is reached. The Gated Recurrent Unit (GRU) network \citep{ChVaGuBaBoScBe14} is a slight simplification of the LSTM, but addresses the vanishing gradient problem in a similar manner. Below we present the notation for an LSTM.

For each layer $l = 0, \dots, L_R-1$ and each time step $t = 1,\dots,T_j$:

\begin{align}
\text{forget gate:}\qquad \mathbf{f}^{l+1}_{j,t} &= \sigma\!\left(\mathbf{W}^{l+1}_{hf}\mathbf{h}^{l+1}_{j,t-1} + \mathbf{W}^{l+1}_{xf}\mathbf{h}^{l}_{j,t} + \mathbf{b}^{l+1}_{f} \right), \label{E_LSTM1} \\
\text{input gate:}\qquad \mathbf{i}^{l+1}_{j,t} &= \sigma\!\left(\mathbf{W}^{l+1}_{hi}\mathbf{h}^{l+1}_{j,t-1} + \mathbf{W}^{l+1}_{xi}\mathbf{h}^{l}_{j,t} + \mathbf{b}^{l+1}_{i} \right),  \label{E_LSTM2}\\
\text{candidate cell state:}\qquad \mathbf{g}^{l+1}_{j,t} &= \phi^{l+1}\!\left(\mathbf{W}^{l+1}_{hg}\mathbf{h}^{l+1}_{j,t-1} + \mathbf{W}^{l+1}_{xg}\mathbf{h}^{l}_{j,t} + \mathbf{b}^{l+1}_{g} \right),\label{E_LSTM3} \\
\text{output gate:}\qquad \mathbf{o}^{l+1}_{j,t} &= \sigma\!\left(\mathbf{W}^{l+1}_{ho}\mathbf{h}^{l+1}_{j,t-1} + \mathbf{W}^{l+1}_{xo}\mathbf{h}^{l}_{j,t} + \mathbf{b}^{l+1}_{o} \right), \label{E_LSTM4}\\
\text{cell state:}\qquad \mathbf{c}^{l+1}_{j,t} &= \mathbf{f}^{l+1}_{j,t} \odot \mathbf{c}^{l+1}_{j,t-1} + \mathbf{i}^{l+1}_{j,t} \odot \mathbf{g}^{l+1}_{j,t}, \label{E_LSTM5} \\
\text{hidden state:}\qquad \mathbf{h}^{l+1}_{j,t} &= \mathbf{o}^{l+1}_{j,t} \odot \tanh\!\left(\mathbf{c}^{l+1}_{j,t}\right), \label{E_LSTM6}
\end{align} 
where $\sigma(x)=1/(1+e^{-x})$ is the sigmoid function, $ \mathbf{a} \odot \mathbf{b}$ is the Hadamard product of matrices $\mathbf{a}$ and $\mathbf{b}$, and $\mathbf{W}^{l+1}_{.f}, \mathbf{W}^{l+1}_{.i}, \mathbf{W}^{l+1}_{.g}, \mathbf{W}^{l+1}_{.o}$ are the weight matrices for the forget gate, input gate, candidate cell state and output gate, respectively. Here, $\phi^{l+1}(\cdot)$ denotes a generic activation function. In the LSTM formulation considered in this paper, the hyperbolic tangent function is used for both the candidate cell state and the hidden state update.

Equations~\eqref{E_LSTM1}--\eqref{E_LSTM6} describe how the LSTM processes an individual claim’s transactional history by selectively retaining, updating and exposing information over time. At each transaction time~$t$, the input vector $\mathbf{h}^{l}_{j,t}$ summarises the contemporaneous state of observation~$j$, including calendar and development timing, cumulative payments and, when available, the case estimate. The weight matrices $\mathbf{W}^{l+1}_{x\cdot}$ govern how these current transactional features influence each gate, while the matrices $\mathbf{W}^{l+1}_{h\cdot}$ determine how information from the previous hidden state $\mathbf{h}^{l+1}_{j,t-1}$ propagates forward through time.

The forget gate $\mathbf{f}^{l+1}_{j,t}$ in~\eqref{E_LSTM1} controls which components of the previous cell state $\mathbf{c}^{l+1}_{j,t-1}$ are retained, allowing the network to learn when historical information should lose relevance, for example as a claim transitions from early investigation to settlement. The input gate $\mathbf{i}^{l+1}_{j,t}$ in~\eqref{E_LSTM2} regulates how much new information enters the memory through the candidate state $\mathbf{g}^{l+1}_{j,t}$ defined in~\eqref{E_LSTM3}, enabling the model to respond more strongly to structurally informative events such as large payments or major case estimate revisions. The resulting cell state $\mathbf{c}^{l+1}_{j,t}$ in~\eqref{E_LSTM5} therefore acts as a latent summary of the claim’s development, combining selectively retained historical information with newly incorporated signals. Finally, the output gate $\mathbf{o}^{l+1}_{j,t}$ in~\eqref{E_LSTM4} determines how much of this latent memory is exposed via the hidden state $\mathbf{h}^{l+1}_{j,t}$ in~\eqref{E_LSTM6}, which is passed to subsequent layers and ultimately influences the predicted ultimate claim size.

While the LSTM and GRU models have theoretical advantages over a vanilla RNN, in practice these do not always manifest. Therefore, when building an RNN, LSTM or GRU, all three variants should be tested. In this paper, we will only consider LSTMs, as \citet*{Lam24} suggests that it outperforms both the vanilla RNN and GRU in this context.

\subsection{Comparison of FNNs and RNNs}

The distinction between feed-forward and recurrent architectures can be made precise by comparing equation~\eqref{E_FNN} with equation~\eqref{E_RNN} and the LSTM formulation in equations~\eqref{E_LSTM1}--\eqref{E_LSTM6}. In the feed-forward setting, each observation is mapped to an output through a fixed sequence of transformations that depend only on the inputs presented at the prediction time. Any temporal structure in a claim’s development must therefore be conveyed indirectly, either through engineered summary statistics or through static representations of past behaviour. By contrast, the recurrent formulations explicitly model claim development as a sequential process, in which information from earlier transactions influences later representations through recurrent connections and a persistent latent state.

From a reserving perspective, this difference is substantive rather than merely architectural. Equations~\eqref{E_LSTM1}--\eqref{E_LSTM6} embed the assumption that claim histories contain ordered information whose relevance evolves over time, whereas equation~\eqref{E_FNN} processes the input as a static vector once presented to the network, so that any temporal ordering, recency or persistence must be encoded ex ante through feature engineering rather than learned through the model structure. This raises a natural empirical question: whether the structural advantages of recurrent networks translate into improved predictive performance when full transactional histories are available, or whether carefully constructed summary statistics are sufficient to capture the information required for accurate reserving. This forms the basis of our first research question.

Although feed-forward networks are, in principle, universal function approximators \citep{Cyb89}, in practice it can be difficult for them to learn complex temporal relationships efficiently \citep{Elm90}. In time series applications, recurrent neural networks are therefore more commonly used, as they explicitly encode sequential dependence through their architecture. By contrast, models such as standard feed-forward networks must learn these relationships implicitly from the data, without any structural guidance. As a result, a substantial portion of their training effort may be devoted to identifying temporal patterns that are already hard-coded into the recurrent formulation.

This distinction becomes particularly relevant when considering individual claim histories. While it is possible for feed-forward networks to handle time series inputs and potentially learn temporal relationships, doing so is more challenging from both an implementation and a structural perspective. The formulae in Section~\ref{sec:rnn} illustrate how time series inputs are naturally accommodated within an RNN. By contrast, replicating this behaviour in a feed-forward network would require defining the input vector as $\mathbf{x}_j = (x_{j,1}, x_{j,2}, \ldots, x_{j,T_j})^T$. However, while an RNN can directly handle sequences of varying length, feed-forward networks cannot, as the number of weights depends on the input dimension $p^{(0)}$, which in this case corresponds to the sequence length.

To address this limitation, sequences must be standardised to a fixed length $T$ through padding (for shorter sequences) and truncation (for longer ones). Further challenges arise when multiple time series are observed simultaneously. While these are explicitly included in the definition of $\mathbf{x}_{j,t}$ for recurrent networks, a feed-forward network requires these series to be flattened into a one-dimensional input vector, for example
\[
\mathbf{x}_j = (x_{j,1,1}, x_{j,1,2}, \ldots, x_{j,1,T}, x_{j,2,1}, \ldots, x_{j,p^{(0)},T})^T \in \mathbb{R}^{T \cdot p^{(0)}}.
\]
In this representation, $T$ denotes the maximum sequence length after padding and truncation, and $p^{(0)}$ the number of concurrent time series. The network must then learn temporal dependencies within and across each series, as well as implicitly infer where one series ends and another begins. These additional requirements increase model complexity and training effort, without any guarantee that a feed-forward network will perform as well as, or better than, a recurrent alternative.

\subsection{Inputs} \label{sec:inputs}

The LSTM contains two different types of inputs; refer back to Figure \ref{fig:LSTM Architecture}. The first is a set of time series that are fed directly into the recurrent layers, containing all transactions that occur before the prediction time. These include the calendar time of each transaction, the time of each transaction since notification of the claim, the cumulative payments made at each transaction time and the case estimate at each transaction time.

Similar to how a claim's payment history can be thought of as a time series, the history of a claim's case estimate developments also forms a time series. Since the LSTM is already equipped to handle (and indeed benefits from analysing) time series data, it makes sense to also include the case estimate data in our model, if available. The case estimate histories, if nothing else, would at least provide a starting point for predictions made before the first payment on a claim. Additionally, they may encode other useful information about a claim, such as the extent of injury to the claimant or any updates through the legal system, if applicable. This is not something that is explored by \citet*{Kuo20}, \citet*{ChBeCoCo23} or \citet*{ScSc25}, as they use individual loss data with LSTM units but do not use case estimate histories. We denote the model that includes case estimate histories as the LSTM+, while the model without this data will be referred to as the LSTM.

The second set of inputs are combined with the outputs of the LSTM layers and fed directly into the feed-forward layers. These are `static' inputs, as they only take a single value per observation. These include the prediction quarter, accident quarter, and the default covariates simulated in the SPLICE package \citep*{AvTaWaWo21,AvTaWa23}.

\citet*{Lam24} experimented with some `derived' features in their RNN models. These were inputs that could be derived from the information already presented to the network. For example, the `multiplier' and `transaction type' columns from the raw SPLICE data were passed into the recurrent layers, while static outputs such as the mean, variance and maximum of payments to date, as well as number and maximum of case estimate revisions were passed directly into the feed-forward layers. Ultimately it was found that these did not improve model performance. Consequently, we will not revisit these derived features here.

Like the LSTM, the FNN will have a distinction between the model that includes case estimate data (FNN+) and the one without (FNN). Both models will include summarised payment data, as well as the accident and development quarters, and SPLICE covariates.

The summarised data (``payment summaries'' in Figure \ref{fig:FNN Architecture}) consists of the number of payments, mean of payments, coefficient of variation of payments and the maximum payment before the prediction time. The case estimate summary data consists of the sum of the payments to date and the latest case estimate as at the time of the input observation, the number of revisions, mean revision amount, largest revision (in absolute terms), total value of revisions (sum of absolute value of revisions) and the proportion of revisions that were upward before the prediction time. Note that all inputs to the FNN and FNN+ are static.

Visual summaries of these models are contained in Figures \ref{fig:FNN Architecture} and \ref{fig:LSTM Architecture}. The orange nodes are fed as time series inputs into the recurrent layers in the LSTM, whereas in the FNN they are summarised into static inputs before being fed directly into the feed-forward layers.

\subsection{Output} \label{output}

There are currently two major schools of thought regarding what should be predicted by an individual claims model: the first being the ultimate size of each claim at finalisation \citep*{DuPi19, LoMiTh19}, while the second is to predict what changes will occur in the next period (whether payments or incurred claim amounts) and then sequentially use these predictions to generate a full history of changes over the next $v$ periods \citep*{Kuo20, ChBeCoCo23, ScSc25, Gab21, DeWu20}. We have chosen the former approach to reduce the risk of compounding forecast errors across successive quarters. The main caveat with our chosen approach is that the data needs to be carefully split into training, validation and test sets without introducing data leakage; see Section \ref{train-test-split} for further discussion.

Like most insurance-related data, the distribution of ultimate claim sizes in this dataset is heavily skewed. Therefore, using $\log(\text{claim size})$ as the target output aims to reduce the heaviness of the right tail and improve the efficiency of the training process. Normalisation of the target output was necessary for the neural networks to learn effectively. Similar observations were made by \citet*{AMAvTaWo22}.

Another important modelling choice is whether to produce deterministic or stochastic estimates of the individual claim sizes. In a neural network context, a deterministic model exists when the output of the model is a point estimate $y_j \in \mathbb{R}$. Stochastic models, however, typically produce multiple outputs $\mathbf{y}_j \in \mathbb{R}^{p^{(L_F)}}$ that represent the parameters of a probability distribution of the ultimate claim size for observation $j$. See \citet*{AMAvTaWo22} for an example of a neural network with stochastic properties. Alternatively, one could produce a distribution of predictions by retraining a deterministic model under multiple different seeds \citep{Gab21} or by bootstrapping \citep*{GaRiWu19}. However, the runtime to produce predictions will scale linearly by the number of times each model is retrained.

While stochastic models provide a greater insight into the uncertainty surrounding a given estimate \citep*{EnVe02} and are useful for determining appropriate risk margins, we focus here on central estimates only for simplicity.

\subsubsection{Prediction error bias and correction}

Since the models predict on a (normalised) log scale under the Mean Squared Error (MSE) loss function, the models produce predictions of the median instead of the mean when converting to a dollar scale. Therefore, bias correction is necessary to produce dollar scale estimates of the mean instead of the median. We employ a non-parametric bias correction \citep{Dua83}:

\begin{equation}
    \hat{Y}_j = \exp(\hat{y}_j) \cdot \hat{b}, \qquad \hat{b} = \frac{1}{n} \sum_{j=1}^n \exp(y_j - \hat{y}_j),
\end{equation} where
\begin{itemize}
    \item $\hat{Y}_j$ is the model's prediction of observation $j$ on a dollar scale;
    \item $\hat{y}_j$ is the model's prediction of observation $j$ on a log-dollar scale;
    \item $y_j$ is the actual value of observation $j$ on a log-dollar scale.
\end{itemize} Bias correction is only applied to a model's predictions during the evaluation of the validation and test sets after hyperparameter tuning has been conducted. In other words, the bias correction has no influence on the hyperparameters that are selected, as this may lead to a selection of a suboptimal hyperparameter combination which may distort the model's ability to generalise to unseen data in the test set. For each test dataset $d_\text{test}$, the bias correction factor $\hat{b}_d$ is calculated using the model predictions for all observations in the corresponding validation set $d_\text{val}$. When presenting the results from the validation set used for hyperparameter tuning, it is calculated using, and applied to, the observations from that validation set after the completion of the hyperparameter tuning process.

Even though the correction factor is quite blunt, it was crucial for the performance of each of our models at an aggregate level. Although not explored here, it could potentially be extended to have a different correction factor for each partition of the data. For example, a separate bias correction factor for claims that arise in the first 10 accident quarters relative to those that are incurred later on.

For further discussions about bias as well as other potential corrections, we direct the reader to \citet*{Wut20b}, \citet*{DeChTr21} and \citet*{Wut22}.

\section{Training methodology} \label{Training Methodology}

\subsection{Data sets}

The data used for this project was simulated in R using the SynthETIC \citep*{R-SynthETIC,AvTaWaWo21} and SPLICE \citep*{R-SPLICE,AvTaWa23} libraries. SynthETIC is a modular package that simulates claim histories, while SPLICE is an extension to simulate case estimates for each claim history. This particular simulator is designed to mimic the experience of a real motor bodily injury portfolio, under the simulator's default parameterisation.

Both claim payments and case estimate revisions are simulated in continuous time and at an individual claim level. Revisions are classified as either `major' or `minor'. As the names suggest, major revisions tend to be larger in magnitude and less frequent than minor ones. The former can arise due to the emergence of significant information, such as the outcome of a large court case, whereas the latter will usually occur due to finer adjustments in the details of a claim. Both can occur concurrently with a payment.

The simulation process for both payment and revision histories are split into sequential modules, each of which can be adjusted by the user. As a rough guideline, the first quantity to be simulated is the occurrence time for each claim. This is followed by the size of each claim, as well as the reporting and settlement dates. From there, the frequency, size and timing of payments are simulated. Finally, the frequency, timing and size of revisions are simulated. This ordering is carefully chosen due to the dependence structure between modules. For example, the timing of revisions is dependent on the number of revisions, and hence the number of revisions is simulated before the timing of them. More detailed information can be found in their respective papers.

\begin{figure}[!htb]
\begin{tikzpicture}
\begin{axis}[xlabel = Calendar Quarter,
             ylabel = Claim Number,
             xtick pos = left,
             ytick pos = left,
             xtick distance = 10,
             ytick distance = 1,
             grid=major,
             width = \textwidth,
             height = 4cm,
             legend style={
        at={(0.5,-0.5)}, 
        anchor=north,   
        legend columns=3 
    }]

\addplot+ [only marks, 
           mark=o, mark size=3pt,
           black, 
           mark options={fill=white}] coordinates{(2, 1) (15, 2) (6, 3) (45, 4)};
\addlegendentry{Notification}

\addplot+ [only marks, 
           mark=square*, 
           black, 
           mark options={fill=black}] 
           coordinates{(4, 1) (6, 1) 
                       (17, 2) (19, 2) (25, 2) (30, 2) 
                       (10, 3) (20, 3) (22, 3) (45, 3) (52, 3) (57, 3)
                       (47, 4) (50, 4) (57, 4)}; 
\addlegendentry{Payment and/or case estimate revision}

\addplot+ [only marks, 
           mark=triangle*, mark size=3.2pt, 
           black,
           mark options = {fill=black}] 
           coordinates{(7, 1) (38, 2) (60, 3) (60, 4)}; 
\addlegendentry{Settlement}

\addplot[domain=2:7, forget plot] {1}; 
\addplot[domain=15:38, forget plot] {2}; 
\addplot[domain=6:60, forget plot] {3}; 
\addplot[domain=45:60, forget plot] {4}; 

\draw[very thick, dashed] (axis cs:36,0) -- (axis cs:36,5);

\draw[very thick, dashed] (axis cs:40,0) -- (axis cs:40,5);

\end{axis}
\end{tikzpicture}
\caption{Chart illustrating the claim lifecycle. Claims are reported (circles), then payments and/or case estimate revisions occur (squares), and finally the claims are settled upon their final payments (triangles). The dashed lines represent the boundaries of the validation and test sets. All claims that are finalised before the dashed lines will be in the training set. Claims that are finalised within the dashed lines will either be in the training or validation sets. Claims that are finalised after the dashed lines will be in the test set, provided they are reported before the second dashed line. Claims that are reported after both dashed lines are IBNR claims.}
\label{fig:claim-lifecycle}

\begin{center}
\begin{tikzpicture}
\begin{axis}[
    width=12cm,
    height=5cm,
    xmin=-0.05, xmax=7.1,
    ymin=0, ymax=8500,
    xlabel={Calendar Quarter},
    ylabel={Amount (\$)},
    title={Development history of claim \#1},
    xtick={0,1,2,3,4,5,6,7},
    ytick={0,2000,4000,6000,8000},
    scaled y ticks=false,
    yticklabel style={
        /pgf/number format/fixed,
        /pgf/number format/1000 sep={,}
    },
    xtick pos = left,
    ytick pos = left,
    ymajorgrids,
    xmajorgrids=false,
    xminorgrids=false,
    legend style={
        at={(0.5,-0.35)},
        anchor=north,
        legend columns=4,
        /tikz/every even column/.append style={column sep=0.5cm,
        nodes={scale=0.8, transform shape}}
    }
]

\pgfplotsset{
    major revision legend/.style={
        legend image code/.code={
            \draw[black,dashed,thick]
                (0cm,-0.2cm) -- (0cm,0.2cm);
        }
    },
    minor revision legend/.style={
        legend image code/.code={
            \draw[gray,dotted,thick]
                (0cm,-0.2cm) -- (0cm,0.2cm);
        }
    }
}

\addlegendimage{gray,line width=1.0pt}
\addlegendentry{Paid}

\addlegendimage{black,very thick}
\addlegendentry{Incurred}

\addlegendimage{major revision legend}
\addlegendentry{Major revision}

\addlegendimage{minor revision legend}
\addlegendentry{Minor revision}

\addplot[
    gray,
    line width = 1.0pt,
    const plot
]
coordinates{
(2.3,0)
(3.5,2500)
(5.7,5800)
(6.8,7300)
};

\addplot[
    black,
    very thick,
    const plot
]
coordinates{
(2.3,6500)
(5.7,7300)
(6.8,7300)
};

\addplot[
    black,
    dashed,
    thick
] coordinates {(2.3,0) (2.3,10000)};


\addplot[
    gray,
    dotted,
    thick
] coordinates {(5.7,0) (5.7,10000)};

\end{axis}
\end{tikzpicture}
\end{center}
\caption{Example of a claim's development history from notification to settlement. Illustrates claim \#1 from Figure \ref{fig:claim-lifecycle}. Notification of a claim and the consequent creation of a case estimate is considered to be a major case estimate revision.}
\label{fig:claim-development}
\end{figure}

As mentioned in \citet*{AvTaWaWo21,AvTaWa23}, using simulated data serves as a useful tool to test whether a model is able to identify known relationships that have been intentionally included in the data. Furthermore, given the scarcity of large datasets (especially with case estimate data), one often has no choice but to rely on simulated data.

The models discussed in this paper were tested on fifty ``complexity 5'' datasets from SPLICE. These datasets significantly breach the cross-classified assumptions underlying methods such as the chain ladder and cross-classified GLM variants, for example through calendar and occurrence period superimposed inflation that each vary by claim size. Complexity 5 is the highest level of complexity available by default in the simulator. Results under five simpler ``complexity 2'' datasets, which deviate slightly from a chain ladder environment via a dependence of notification delay on claim size, can be found in the supplementary material with similar conclusions. Each dataset contains roughly 30,000 claims with around 250,000 transactions. Time is expressed in quarters. 

Figures \ref{fig:claim-lifecycle} and \ref{fig:claim-development} illustrate the claim lifecycle and development history from claim notification to settlement. Claims are closed immediately upon receipt of the final payment. Once a claim is closed, it cannot be re-opened. For additional visualisations of the raw simulated data, please refer to Table C.1 in \citet*{AvTaWa23}.

Ultimately, we want to build a model that is able to provide a robust estimate of the ultimate claim size at \textbf{any} point during the development of a claim. That is, we want the ability to produce estimates at various points throughout a claim's lifetime. Therefore, we decided to observe (and input) data as of the end of each calendar quarter in which a given claim is still open. More formally, if a claim is open between $t_1$ and $t_2$ ($t_1 < t_2$), then a prediction will be produced at times $\lfloor t_1 \rfloor + 1, \lfloor t_1 \rfloor + 2, ..., \lfloor t_2 \rfloor$. This means there will be a total of $\lfloor t_2 \rfloor - \lfloor t_1 \rfloor$ predictions produced for this claim, which will be used for training the model. Note that these observations are treated as independent during training, even though they are not. We ignore this limitation in this paper, in line with the existing literature.

To achieve this, we duplicate each claim for each prediction time, including all available information up to that prediction time. Each of these duplicates will be treated as one observation for our model. Thus, claims that are reported and settled within a single calendar quarter are excluded from the analysis, while the remaining claims are converted into at least one - and mostly more than one - observation. 

For example, the claim in Figure \ref{fig:claim-development} was notified during calendar quarter 2 and settled in calendar quarter 6. This means there are 4 calendar quarters where this claim is open and has already been notified to the insurer. Therefore, the models should generate predictions for this claim at the beginning of calendar quarters 3, 4, 5 and 6. The observation at quarter 3 would only contain information about the notification of the claim, the observations at quarters 4 and 5 would also contain the first payment transaction while the observation at quarter 6 would contain the information about the minor revision that coincides with a payment during quarter 5. Although the claim development pattern in this example is deliberately simple for illustration purposes, most claims in the complexity 5 datasets feature much more complex patterns in their development history. As such, there are likely to be significant differences between the resulting observations for a single claim, more so than the observations related to the claim in Figure \ref{fig:claim-development}.

\subsection{Train-test split} \label{train-test-split}

\subsubsection{Avoidance of data leakage}

In standard machine learning applications \citep{JaWiHaTiTa23}, the decision of train-test split is often straightforward. However, the nature of insurance claims data possesses particular structure that needs special care where, if ignored, can manifest in issues such as accidental ``data leakage” where information is present (either implicitly or explicitly) in the training set that should not be available during training. As such, in this paper, we will initially split claims based on their finalisation time. 

The training set contains all observations from claims that are finalised before calendar quarter 36, the validation set is for those that are finalised between calendar quarters 36 and 40, while the test set contains those finalised after calendar quarter 40. Note that calendar quarter 40 is the final accident quarter and hence can be viewed as a valuation date. This means that all claims are incurred by calendar quarter 40, but some may not be reported until after this date. These are Incurred But Not Reported (IBNR) claims and are therefore not considered by our models when constructing a reserve of Reported But Not Settled (RBNS) claims at the valuation date. Applying this splitting method to the claims in Figure \ref{fig:claim-lifecycle}, claim \#1 would be in the training set, claim \#2 would be in the validation set, claim \#3 would be in the test set and claim \#4 would be considered IBNR.

If one wanted to estimate the cost of IBNR claims, a chain ladder-based approach could be used, similar to \citet*{Wut18b} and \citet*{DeMo19}. An alternative approach that is more aligned to the micro-level nature of a reported claims model would be to utilise granular policy and exposure-level information to estimate IBNR claim frequency and severity, similar to \citet*{BaRo19}. The IBNR estimate can then be combined with the RBNS reserve to produce an overall outstanding claims reserve.

Importantly, this splitting method ensures that the model is trained and validated solely on past data and hence there is no data leakage from the future. If this was not the case, the models would learn future inflation patterns and shocks, potential regulatory changes, or other information that should not be known at the valuation date. Consequently, the forecasting performance of a model would be overstated.

As alluded to in Section \ref{output}, these considerations are overlooked by \citet*{DuPi19}. Of the papers that do not contain data leakage in their train-test splitting, there are additional concerns elsewhere in their treatment of the data, namely the treatment of large claims. \citet*{ChBeCoCo23} create a separate model for large individual payments based on Extreme Value Theory (EVT), while \citet*{Poo19} enforces a cap on the target claim sizes and \citet*{ScSc25} remove all claims initially reported as `large' from their dataset. We argue that large claims should \textbf{not} be pre-processed differently to smaller claims, as the primary purpose of an individual claims model should be to accurately identify and forecast all claims, especially larger ones due to their relative importance for producing a reserve estimate.

\citet*{Kuo20} is the only paper that, to our knowledge, uses an LSTM in an individual claims forecasting setting without treating large claims or containing any data leakage in the train-test splits. However, the reserve forecasts produced by this model were subject to greater prediction error than those generated by the Chain Ladder. \citet*{Gab21} instead presents a feed-forward neural network that is free from data leakage or major simplifications, however uses the `sequential' prediction approach. Additionally, both papers use datasets generated from the individual claims history simulator in \citet*{GaWu18}, which is a sufficiently different environment from what is generated by SynthETIC \citep*{AvTaWaWo21} and SPLICE \citep*{AvTaWa23}. Table 5.1 in \citet*{AvTaWaWo21} presents a comparison of the simulators. Consequently, the results presented in Section \ref{Results} will not be directly comparable to those presented in \citet*{Kuo20} or \citet*{GaWu18}.

\subsubsection{Methods to mitigate bias}

Our proposed splitting method does introduce a significant challenge, however. If claims are split solely based on their finalisation time, then the training set would be biased to contain claims with relatively short developments while the test set would be biased towards claims with longer developments. This means there is likely to be a significant difference between the types of claims that the model is trained on relative to those it is expected to forecast. In other words, there is a trade-off between teaching a model to extrapolate and ensuring the training set is comprehensive enough for reliable extrapolation. 

Fixed origin and rolling origin cross-validation \citep{Tas00} are two potential mitigants. However, relative to a singular validation set, these methods would increase the runtime costs by a factor equal to the number of windows, or splits. In practice, the number of windows is usually chosen to be between five and ten, which could easily increase the runtime of a model by days or weeks.

Another alternative, as employed by \citet*{LoMiTh19}, is to use a Weighted Mean Squared Error (WMSE) loss function to give greater weight to training observations with longer developments. To determine the weights, they make use of the Kaplan-Meier estimator and assign a weight of 0 to claims that have not been finalised. This therefore avoids any data leakage issues. While this approach will increase the weights for claims that are close to finalisation, we instead want to emphasise the early development observations for claims that have relatively long settlement times.

Our solution is to randomly move 20\% of the validation set into the training set. This means that claim \#2 in Figure \ref{fig:claim-lifecycle} could appear in either the training or validation set, depending on the random seed. It will ensure that some slightly longer development claims will be contained in the training set and that the model will observe all accident quarters in the training data. The training set will still be biased towards claims that finalise before quarter 36, while the validation set will still be comprised solely of claims that finalise between quarters 36 and 40. The final train-validation-test split at an observation level is roughly 48\%-8\%-44\% for each simulated dataset. For each simulated dataset, a random seed is used when randomly moving these additional claims into the respective training set.

\subsection{Metrics and benchmarks} \label{sec:Metrics}

\subsubsection{Individual claims metrics}

We are first interested in developing a measure that teaches us how ``often'' a model's prediction ``beats'' that of another (including case estimates). We denote the `M1vsM2' metric as one such measure of performance:
\begin{equation} \label{E_vsCE}
\text{M1vsM2} = \sum_{j=1}^{n}{w_j \cdot I\{|\hat{Y}_{j, M1} - Y_j| < |\hat{Y}_{j, M2} - Y_j|\}},
\end{equation}  where $Y_j$ is the ultimate claim size for observation $j$, $\hat{Y}_j$ is the model prediction of the ultimate claim size for observation $j$, $I\{.\}$ is an indicator function and $w_j$ is an appropriate weight. All quantities are on a dollar scale.

When $w_j=1/n$, \eqref{E_vsCE} simply calculates how often the prediction of M1 was closer (as compared to that of M2) to the actual ultimate claim size. This presents a number of issues. Because \eqref{E_vsCE} does not consider the scale of the deviations to $Y_j$, the metric may end up being dominated by smaller claims. Furthermore, we are naturally more interested in performance on those claims that have a large amount outstanding (rather than ultimate claim size, or cumulative payments thus far). If M2 are `case estimates', $w_j=1/n$ presents yet another issue. Case estimates have a tendency to converge to the true amount outstanding as the claim approaches finalisation. Once a case estimate equals the true outstanding cost, or gets close enough, it becomes strictly, or near, impossible for the model to outperform it. Additionally, for smaller and less complex claims, we would expect the case estimates to be relatively accurate from notification. In view of those observations, consider the following two possible alternative weights,
\begin{equation}
w_j^{\text{S}} = \frac{Y_j}{\sum_{k=1}^{n}{Y_k}}, \quad \text{ or } w_j^\text{O} = \frac{Y_j-P_j}{\sum_{k=1}^{n}{(Y_k-P_k)}},
\end{equation}
where $P_j$ denotes the cumulative payments made for the claim in observation $j$ up to the prediction time. $w_j^\text{S}$ is defined as a weight proportional to the actual ultimate claim size of each observation, whereas $w_j^\text{O}$ is a weight proportional to the actual amount outstanding (i.e. the actual claim size less payments made to date) for each observation. Both weighting structures emphasise larger claims, reflecting the relative importance to the insurer of these claims. Weight $w_j^\text{O}$ goes one step further by emphasising observations that are earlier in their development, as they will have a relatively larger amount outstanding compared to later in their development. Therefore, for the remainder of this paper, any references to `M1vsM2' type metrics will be using $w_j^\text{O}$ (unless otherwise specified). Furthermore, in the case of comparisons with case estimates, we will write $\text{vsCE}_\text{O}$ and will refer to a slightly modified version of Equation \eqref{E_vsCE}: 

\begin{equation}
    \text{M1vsCE}_\text{O} = \sum_{j=1}^{n}{w_j^\text{O} \cdot I\{|\hat{Y}_{j, M1} - Y_j| < |(\tilde{Y}_{j} + P_j) - Y_j|\}},
\end{equation} where $\tilde{Y}_j$ is the case estimate for observation $j$. $\text{M1vsCE}_\text{S}$ and $\text{M1vsCE}$ are similarly defined using weights $w_j^\text{S}$ and $w_j=1/n$, respectively. When considering case estimates as a `model' in this manner, we will refer to, for each observation that will be used to calculate one of these metrics, the latest available case estimate at the time of the observation. For completeness, we have included results for the cases where $w_j=1/n$ and $w_j^{\text{S}}$ in the supplementary material. The `vs' metrics displayed in this paper and the supplementary material are grouped by weight, as listed in Table~\ref{tab:vs-metrics}.

\begin{table}[!htb]
    \centering
    \begin{tabular}{@{}cp{0.85\textwidth}@{}}
    \toprule
    Weight & `vs' metrics \\ \midrule
    $w_j^\text{O}$ & LSTM+vsFNN+, FNN+vsFNN, $\text{LSTM+vsCE}_\text{O}$, $\text{LSTMvsCE}_\text{O}$, $\text{FNN+vsCE}_\text{O}$, $\text{FNNvsCE}_\text{O}$ \\
    $w_j^\text{S}$ & $\text{LSTM+vsCE}_\text{S}$, $\text{LSTMvsCE}_\text{S}$, $\text{FNN+vsCE}_\text{S}$, $\text{FNNvsCE}_\text{S}$ \\
    $w_j=1/n$ & LSTM+vsCE, LSTMvsCE, FNN+vsCE, FNNvsCE \\
    \bottomrule
    \end{tabular}
    \caption{The `vs' metrics displayed in the paper and the supplementary material, grouped by the weight used.}
    \label{tab:vs-metrics}
\end{table}

If the M1vsM2 metric produces a figure close to 80\%, this would suggest that after weighting for the size of the outstanding amounts owing on each claim, M1 will produce tighter individual forecasts at a greater frequency relative to M2. However, it still does not express the degree of relative accuracy of the individual forecasts. As a result, this metric should not be considered in isolation, but rather in combination with the other metrics.

To complement the above metrics, we also examine the Mean Absolute Logarithmic Error (MALE)
\begin{equation}
    \text{MALE} = \frac{1}{n} \sum_{j=1}^{n}{\left|\log(\hat{Y}_j-P_j) - \log(Y_j-P_j)\right|} = \frac{1}{n} \sum_{j=1}^{n}{\left|\log\left(\frac{\hat{Y}_j-P_j}{Y_j-P_j}\right)\right|},
\end{equation}
and Mean Squared Logarithmic Error (MSLE)
\begin{equation}
    \text{MSLE} = \frac{1}{n} \sum_{j=1}^{n}{\left(\log(\hat{Y}_j-P_j) - \log(Y_j-P_j)\right)^2} = \frac{1}{n} \sum_{j=1}^{n}{\left(\log\left(\frac{\hat{Y}_j-P_j}{Y_j-P_j}\right)\right)^2}.
\end{equation} 
By construction, $\log(\hat{Y}_j-P_j)-\log(Y_j-P_j)$ can be interpreted as an approximate error rate between the model's prediction of the amount outstanding and the true amount outstanding for observation $j$. Therefore, the MALE can be interpreted as an average error rate, and the MSLE as the average of the squared error rates. Like all error functions, a smaller value indicates stronger model performance. 

One potential issue with the above metrics, however, is the possibility of negative model predictions of the amounts outstanding. This would lead to an error when taking the logarithm of the forecasted amount outstanding. Additionally, if the forecasted amount outstanding is positive but close to \$0, the application of the logarithm would lead to an explosion in the error term. To mitigate this, if $\hat{Y}_j - P_j < 1$ for observation $j$, we replace its error contributions of $|\log(\hat{Y}_j-P_j) - \log(Y_j-P_j)|$ and $(\log(\hat{Y}_j-P_j) - \log(Y_j-P_j))^2$ with $|\log(Y_j-P_j)|$ and $(\log(Y_j-P_j))^2$ for the MALE and MSLE, respectively. That way, these models will be penalised heavily from an evaluation standpoint whenever a negative prediction of the amount outstanding is produced. Note that this approach is possible in this environment as there are no recoveries, and hence no negative amounts outstanding, at any point in a claim's development. This assumption is unlikely to hold in practice and would require alternative workarounds.

\subsubsection{Aggregate portfolio metrics}

Even though these models forecast individual claim amounts, eventually these will be aggregated for the purposes of constructing a reserve. Therefore, we also examine the aggregate forecasts produced by the models and case estimates at the valuation date. Our chosen metric is the (RBNS) reserve error:

\begin{equation}
    \text{ResErr}_m = \frac{\sum_{j=1}^{n}{(\hat{Y}_{j,m}-P_j)}}{\sum_{j=1}^{n}{(Y_j-P_j)}} - 1
    \label{eq:ResErr}
\end{equation}  for model $m$. In this case, producing a smaller absolute reserve error is an indication of relative strength.

\subsection{Hyperparameter tuning}

Training a neural network comes with many challenges. Many of these boil down to the structuring of the network and hyperparameter tuning. 

When it comes to the latter, K-Fold Cross Validation \citep{Koh95} is a common method. However, it is not appropriate for forecasting problems. Instead, methods such as Fixed Origin or Rolling Origin Cross-Validation \citep{Tas00} are preferable due to their explicit consideration of time effects. As previously mentioned, however, the main drawback of these methods lies in their increased runtime costs. Consequently, we train the model with each hyperparameter combination once, and evaluate its performance on the validation set.

The other major challenge with hyperparameter tuning is the sheer number of hyperparameters to be tuned. We perform a grid search over a modest selection of hyperparameter combinations where the search range for each hyperparameter was selected through a preliminary trial-and-error process. We tested 24 hyperparameter combinations for all models, noting that there is a significant discrepancy between the runtime of the FNN+ and FNN models relative to the LSTM+ and LSTM. Our criterion for selecting the `best' hyperparameter combination for a model is based on minimising the validation loss, which in this case is the MSE of the natural logarithm of the ultimate claim sizes. During the hyperparameter tuning process, the weights of the neural networks are initially randomised under a set seed of 1. The results presented in Section \ref{Results} are from the models with their selected hyperparameter combination.

Table \ref{fig:hyperparameters} in \ref{app:hyperparameter} summarises the hyperparameter tuning in tabular form. The same values were tested for the `+’ models compared to their `regular’ counterparts. However, each model was tuned separately. We save a discussion of each hyperparameter in \ref{app:hyperparameter}.

\section{Numerical results and discussion} \label{Results}

Throughout this section, model performance is assessed both at the individual-claim level using prediction error metrics and at the portfolio level through aggregate reserve accuracy. All results presented in this section will be across the test sets of fifty datasets, simulated using the highest complexity level available by default in SynthETIC \citep*{R-SynthETIC,AvTaWaWo21} and SPLICE \citep*{R-SPLICE,AvTaWa23}, each generated using a different random seed. 

Multiple datasets and evaluation on a test set are usually infeasible in practice. Evaluation across multiple datasets is uncommon because there is rarely enough data available to split into multiple datasets, though this can be replicated to some extent through bootstrapping. Additionally, results cannot be evaluated on a complete test set as the true claim amounts outstanding are not known at the valuation date and cannot be observed until all claims that were open at the valuation date have been finalised. Nevertheless, we choose to present the results in this manner to generate additional insights into the behaviours of the models that are otherwise unable to be observed. Rolling Origin Cross-Validation \citep{Tas00} or a similar method might aid backtesting and evaluating robustness of models in a more practical setting.

Each model is trained once per dataset, primarily due to runtime considerations. However, the models could instead be trained multiple times per dataset using different random initialisations, similar to what is employed by \citet{Gab21}. The hyperparameter tuning was conducted on the validation set of a separate dataset, again with the same simulation parameterisation. On this separate validation set, tuning was conducted once per model again for runtime considerations. Once a hyperparameter combination has been chosen, that combination is then used for training and evaluating the model on each of the fifty datasets. Please refer to \ref{results validation} for the validation set results.

The reasons for tuning the hyperparameters in this manner are twofold. Firstly, there would be significant runtime challenges if tuning were to be conducted on each of the fifty datasets independently. The other, more theoretical, reason is that it allows us to assess the robustness of each model. Given that each of the datasets is simulated under the same parameterisation, it is expected that the hyperparameter selection between datasets should be similar, if not the same. At the very least, it would result in a slight underperformance of the models compared to if the hyperparameter tuning was performed on each dataset separately.

The results presented in this section will only include observations as at the valuation date. In other words, each model will only produce one prediction per open claim using all available information as at the valuation date. Calendar quarter 40 is chosen as the valuation date as it is the final accident quarter from which claims are simulated. Therefore, this valuation date can be considered to be equivalent to one that would be used by a more conventional `triangular' model.

\subsection{Research Question 1: Models with transactional history versus summarised data} \label{q1}

As a reminder, our first research question asks whether there is a significant difference in performance between a feed-forward network trained solely on summary data compared to a recurrent network trained on full transactional data. The LSTM+ and FNN+ models will be used to answer this question.

\begin{table}[!htb]
    \centering
    \begin{tabular}{@{}crrrrc@{}}
    \toprule
    Model     & ResErr (\%) & MALE          & MSLE  \\ \midrule
    Case Estimates & -33.6 (0.8)    & 0.828 (0.017) & 2.051 (0.066)   \\
    LSTM+     & -6.0 (5.7)     & 0.629 (0.042) & 0.934 (0.218)   \\ 
    FNN+      & -7.7 (6.9)     & 0.550 (0.036) & 0.633 (0.068)   \\ \bottomrule
    \end{tabular}
    \caption{Reserve error, MALE and MSLE metrics for the LSTM+ and FNN+ at the valuation date across 50 datasets. Values are expressed as mean (standard deviation).}
    \label{results table q1}
\end{table}

\begin{table}[!htb]
    \centering
    \begin{tabular}{@{}crrrrc@{}}
    \toprule
    Model          & Case Estimates & FNN+       \\ \midrule
    LSTM+          & 58.8 (2.1)     & 45.4 (4.9) \\ 
    FNN+           & 60.4 (2.7)     & -          \\ \bottomrule
    \end{tabular}
    \caption{`vs table' for the LSTM+ and FNN+. Row $i$, column $j$ displays the `M$i$vsM$j$' metric. Values are expressed as mean (standard deviation).}
    \label{vsCE table q1}
\end{table} 

\begin{table}[!htb]
    \centering
    \begin{tabular}{@{}crrrrc@{}}
    \toprule
                   & $\lvert \text{ResErr} \rvert$ & MALE  & MSLE & $\text{vsCE}_\text{O}$ \\ \midrule
    Test statistic & 359                           & 20    & 15   & 277   \\
    $p$-value      & 0.0065 & $6.5 \times 10^{-13}$ & $2.4 \times 10^{-13}$ & $3.4 \times 10^{-4}$   \\ 
    Win count      & 33                            & 3     & 2    & 15 \\ \bottomrule
    \end{tabular}
    \caption{Wilcoxon two-sided test statistic and $p$-value comparing the LSTM+ and FNN+ for each evaluation metric. The win count represents the number of datasets (out of 50) where the LSTM+ produced a smaller value than the FNN+ for the absolute reserve error, MALE and MSLE metrics, or a larger value for the $\text{vsCE}_\text{O}$ metric.}
    \label{wilcoxon q1}
\end{table}

\noindent Table \ref{results table q1} displays the mean and standard deviation of the metrics defined in Section \ref{sec:Metrics}. Both models perform similarly at an aggregate and an individual claim level. The LSTM+ has a slightly better mean reserve error and tighter variance, while the FNN+ slightly outperforms when looking at both the MALE and MSLE metrics. Table \ref{vsCE table q1} displays the `vs' metrics, with both models asserting similar performances.

Table \ref{wilcoxon q1} instead compares the performance of each model on each individual dataset. It suggests the LSTM+ outperforms the FNN+ when considering the absolute reserve error while the reverse is true for the MALE, MSLE and $\text{vsCE}_\text{O}$ metrics.

\begin{figure}[!htb]
    \centering
    \subfloat[]{
        \includegraphics[height=4.5cm, keepaspectratio]{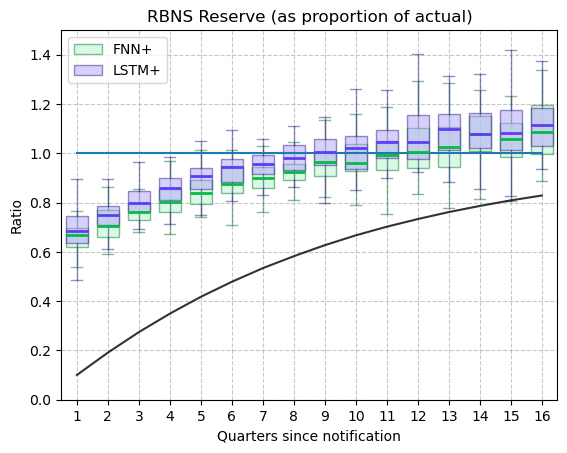}
        \label{fig:LSTMplus_FNNplus_OCLs_dev}
    }
    \subfloat[]{
        \includegraphics[height=4.5cm, keepaspectratio]{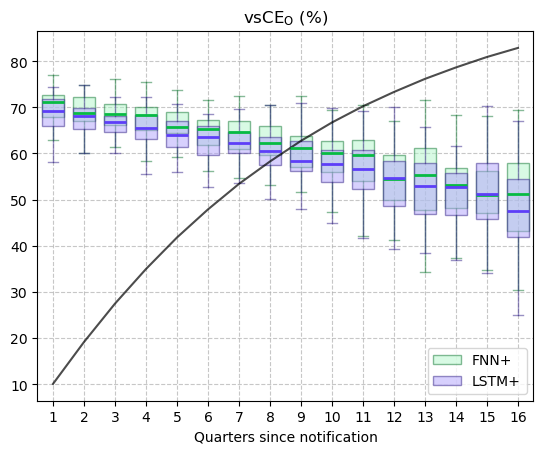}
        \label{fig:LSTMplus_FNNplus_vsCE_OCL_dev}
    }
    \caption{Boxplots showing the aggregate predictions of outstanding claim amounts (expressed as a proportion of the true amounts outstanding) and $\text{vsCE}_\text{O}$ for the LSTM+ and FNN+ at the valuation date, subset by quarters since notification. The black curves represent the proportion of the actual amounts outstanding related to predictions made on or before quarter q. Stronger performance is indicated by values closer to 1 for \protect\subref{fig:LSTMplus_FNNplus_OCLs_dev} and larger values for \protect\subref{fig:LSTMplus_FNNplus_vsCE_OCL_dev}.}
    \label{fig: LSTMplus_FNNplus}
\end{figure} 

\noindent An analysis of the reserve forecasts partitioned by accident quarters and quarters since notification of a claim, in Figure \ref{fig: LSTMplus_FNNplus}, again reveal comparable results. However, one interesting point is that the LSTM+ slightly outperforms the FNN+ at an aggregate level across most quarters since notification. The LSTM+ produces slightly larger forecasts in earlier development periods (quarters 4-10) and slightly smaller forecasts in later development periods (quarters 27-40), leading to a tighter and more balanced reserve forecast. 

\begin{figure}[!htb]
    \centering
    \subfloat[MALE]{
        \includegraphics[height=4.5cm, keepaspectratio]{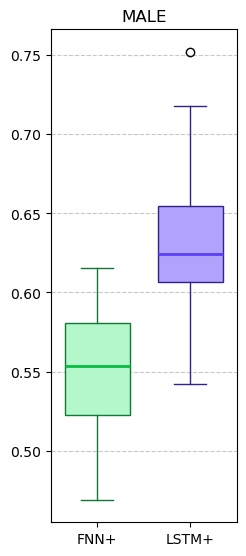}
        \label{fig:LSTMplus_FNNplus_MALE_val}
    }
    \subfloat[MSLE]{
        \includegraphics[height=4.5cm, keepaspectratio]{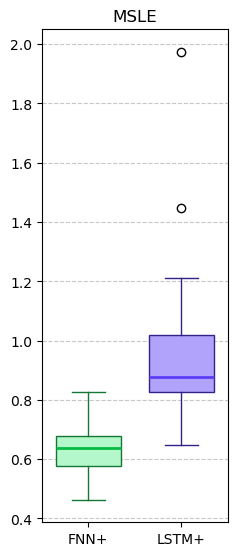}
        \label{fig:LSTMplus_FNNplus_MSLE_val}
    }
    \subfloat[ResErr]{
        \includegraphics[height=4.5cm, keepaspectratio]{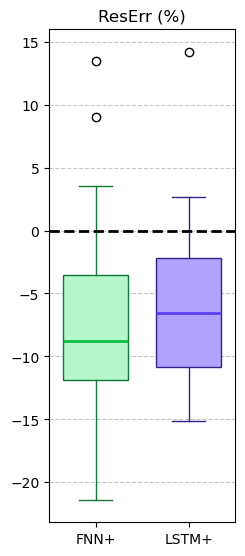}
        \label{fig:LSTMplus_FNNplus_OCLs_val}
    }
    \subfloat[$\text{vsCE}_\text{O}$]{
        \includegraphics[height=4.5cm, keepaspectratio]{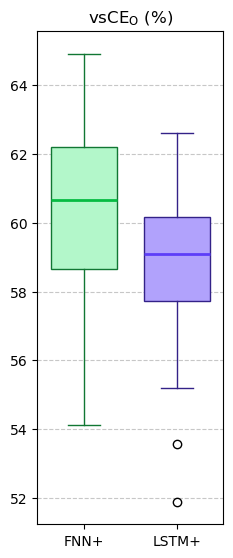}
        \label{fig:LSTMplus_FNNplus_vsCE_OCL_val}
    }

    \caption{Boxplots containing MALE, MSLE, reserve error and $\text{vsCE}_\text{O}$ metrics for the LSTM+ and FNN+ at the valuation date. Smaller values for \protect\subref{fig:LSTMplus_FNNplus_MALE_val} and \protect\subref{fig:LSTMplus_FNNplus_MSLE_val}, values closer to 0 for \protect\subref{fig:LSTMplus_FNNplus_OCLs_val} and larger values for \protect\subref{fig:LSTMplus_FNNplus_vsCE_OCL_val} indicate stronger performance.}
    \label{fig:boxplots-val LSTMplus_FNNplus}
\end{figure} 

\noindent Figure \ref{fig:boxplots-val LSTMplus_FNNplus} provides a visual representation of the statistics in Tables \ref{results table q1} and \ref{vsCE table q1}. In panel (c) it becomes more apparent that the LSTM+ produces smaller reserve errors, as it possesses a median that is closer to 0, and a relatively shorter lower whisker. It also reinforces the notion that the FNN+ outperforms the LSTM+ at an individual claim level (other panels). In particular, the majority of the MALE values produced by the LSTM+ are larger, and therefore worse, than the worst MALE recorded by the FNN+. The MSLE plot leads to similar observations, with the addition of two outliers that further support the claim that the FNN+ outperforms the LSTM+ at an individual level.

Ultimately, while the addition of full transactional data does appear to slightly improve the forecasts in aggregate, it does deteriorate the results at an individual claim level. However, the use of different random seeds to produce the train-test splits could alter the relative performance of each model. From a practical standpoint, the average runtime of the LSTM+ is roughly 50\% longer than that of the FNN+, which could add days to the runtime. The LSTM+ also contains additional modelling complexities, particularly the structuring of the network architecture, that should be considered when evaluating the performance of these models. As such, the LSTM+ in its current architectural form presents limited additional value compared to the FNN+ in this specific environment with sufficiently well-designed features. Again, it is possible that this result may not hold under different circumstances, such as more complex payment and/or case estimate revision patterns or the inclusion of additional covariates in time-series form.

For the equivalent results a more `na\"{\i}ve' train-test split, please see \ref{sec:naive results}. These results demonstrate the degree to which the train-test split influences the perception of model performance. Under the `na\"{\i}ve' split, the performance of the LSTM+ and FNN+ improve dramatically across all metrics. However, this split is not practical, as models cannot be trained on data that will be observed in the future. Please refer to the appendix for further discussion.

\subsection{Research Question 2: Models with and without inclusion of case estimate data} \label{q2}

Research question 2 asks for a comparison between a neural network trained both with and without case estimate data. We will answer this question by comparing the performance of the FNN+ to that of the FNN.

\begin{table}[!htb]
    \centering
    \begin{tabular}{@{}crrrrc@{}}
    \toprule
    Model     & ResErr (\%) & MALE          & MSLE            \\ \midrule
    Case Estimates & -33.6 (0.8)    & 0.828 (0.017) & 2.051 (0.066)   \\
    FNN+      & -7.7 (6.9)     & 0.550 (0.036) & 0.633 (0.068)   \\ 
    FNN       & -19.3 (9.2)    & 0.712 (0.070) & 1.143 (0.400)   \\ \bottomrule
    \end{tabular}
    \caption{Reserve error, MALE and MSLE metrics for the FNN+ and FNN at the valuation date across 50 datasets. Values are expressed as mean (standard deviation).}
    \label{results table q2}
\end{table} 

\noindent Table \ref{results table q2} shows a substantial improvement in the absolute mean reserve error from 19.3\% without the case estimate data to 7.7\% with it. Furthermore, the results at an individual claim level also support the inclusion of case estimate data, as seen through a markedly smaller mean MALE and MSLE for the FNN+ relative to the FNN. Table \ref{vsCE table q2} also supports this, as the FNN+ possesses a higher mean $\text{vsCE}_\text{O}$ with comparable variance, and the FNN+vsFNN being substantially higher than 50\%. Table \ref{wilcoxon q2} reinforces these claims, as it shows that within each of the 50 individual test datasets, the FNN+ always possesses a lower MALE and MSLE, and a higher $\text{vsCE}_\text{O}$, than the FNN. Additionally, the absolute reserve error is smaller for the FNN+ compared to the FNN in 45 out of 50 of the test datasets. The Wilcoxon two-sided test suggests the median reserve error of the FNN+ is significantly lower than that of the FNN.

\begin{table}[!htb]
    \centering
    \begin{tabular}{@{}crrrrc@{}}
    \toprule
    Model          & Case Estimates  & FNN        \\ \midrule
    FNN+           & 60.4 (2.7)      & 56.3 (4.7) \\ 
    FNN            & 50.7 (1.6)      & -          \\ \bottomrule
    \end{tabular}
    \caption{`vs table' for the FNN+ and FNN. Row $i$, column $j$ displays the `M$i$vsM$j$' metric. Values are expressed as mean (standard deviation).}
    \label{vsCE table q2}
\end{table}

\begin{table}[!htb]
    \centering
    \begin{tabular}{@{}crrrrc@{}}
    \toprule
                   & $\lvert \text{ResErr} \rvert$ & MALE  & MSLE & $\text{vsCE}_\text{O}$ \\ \midrule
    Test statistic & 63                            & 0     & 0    & 0   \\
    $p$-value      & $2.5 \times 10^{-10}$ & $1.8 \times 10^{-15}$ & $1.8 \times 10^{-15}$ & $1.8 \times 10^{-15}$   \\ 
    Win count      & 45                            & 50    & 50   & 50  \\ \bottomrule
    \end{tabular}
    \caption{Wilcoxon two-sided test statistic and $p$-value comparing the FNN+ and FNN for each evaluation metric. The win count represents the number of datasets (out of 50) where the FNN+ produced a smaller value than the FNN for the absolute reserve error, MALE and MSLE metrics, or a larger value for the $\text{vsCE}_\text{O}$ metric.}
    \label{wilcoxon q2}
\end{table}

\begin{figure}[!htb]
    \centering
    \subfloat[FNN+]{
        \includegraphics[height=4.5cm, keepaspectratio]{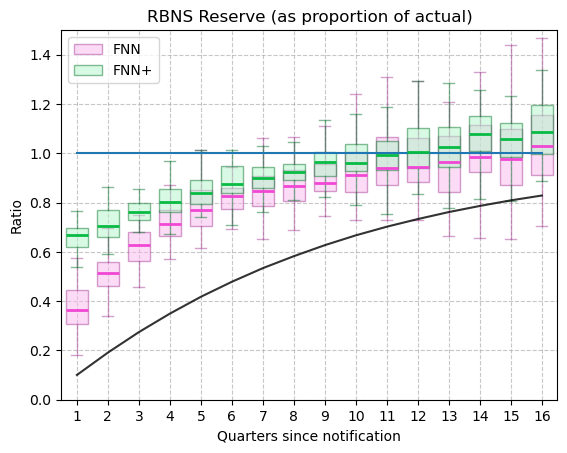}
        \label{fig:FNNplus_FNN_OCLs_dev}
    }
    \subfloat[FNN]{
        \includegraphics[height=4.5cm, keepaspectratio]{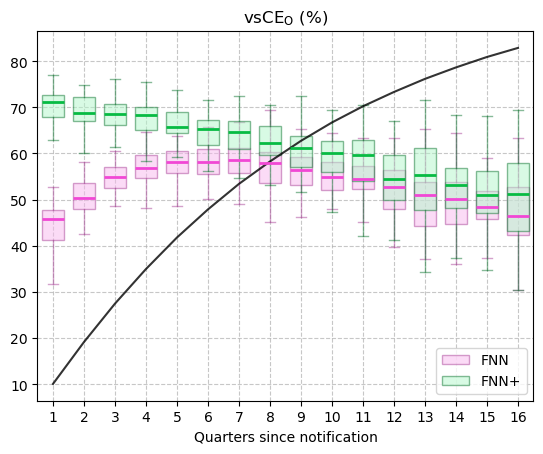}
        \label{fig:FNNplus_FNN_vsCE_OCL_dev}
    }
    \caption{Boxplots showing the aggregate predictions of outstanding claim amounts (expressed as a proportion of the true amounts outstanding) and $\text{vsCE}_\text{O}$ for the FNN+ and FNN at the valuation date, subset by quarters since notification. The black curves represent the proportion of the actual amounts outstanding related to predictions made on or before quarter q. Stronger performance is indicated by values closer to 1 for \protect\subref{fig:FNNplus_FNN_OCLs_dev} and larger values for \protect\subref{fig:FNNplus_FNN_vsCE_OCL_dev}.}
    \label{fig: FNNplus_FNN}
\end{figure}

\noindent Figure \ref{fig: FNNplus_FNN} analyses trends by accident quarter and quarter since notification of a claim. Here it can be seen that the FNN+ produces markedly tighter aggregate forecasts across all periods, whereas the FNN substantially underestimates early in a claim's development and overstates in later periods to compensate. However, the degree of downward biases in the first 10 - and particularly the first 3 - quarters since notification are material enough to deteriorate the overall reserve produced by the FNN.

Therefore, the results presented here suggest that in this specific environment, the case estimate data adds significant value to the model across all claim durations.

\begin{figure}[!htb]
    \centering
    \subfloat[MALE]{
        \includegraphics[height=4.5cm, keepaspectratio]{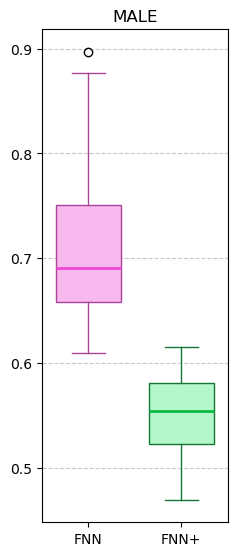}
        \label{fig:FNNplus_FNN_MALE_val}
    }
    \subfloat[MSLE]{
        \includegraphics[height=4.5cm, keepaspectratio]{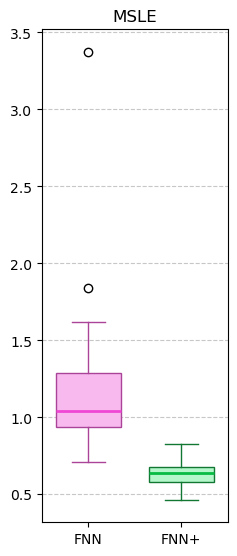}
        \label{fig:FNNplus_FNN_MSLE_val}
    }
    \subfloat[ResErr]{
        \includegraphics[height=4.5cm, keepaspectratio]{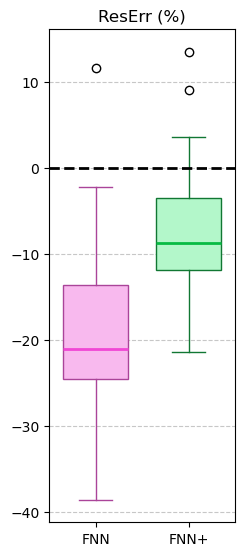}
        \label{fig:FNNplus_FNN_OCLs_val}
    }
    \subfloat[$\text{vsCE}_\text{O}$]{
        \includegraphics[height=4.5cm, keepaspectratio]{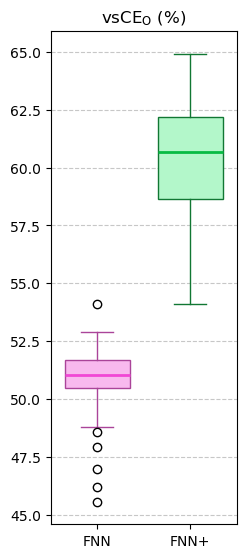}
        \label{fig:FNNplus_FNN_vsCE_OCL_val}
    }

    \caption{Boxplots containing MALE, MSLE, reserve error and $\text{vsCE}_\text{O}$ metrics for the FNN+ and FNN at the valuation date. Smaller values for \protect\subref{fig:FNNplus_FNN_MALE_val} and \protect\subref{fig:FNNplus_FNN_MSLE_val}, values closer to 0 for \protect\subref{fig:FNNplus_FNN_OCLs_val} and larger values for \protect\subref{fig:FNNplus_FNN_vsCE_OCL_val} indicate stronger performance.}
    \label{fig:boxplots-val FNNplus_FNN}
\end{figure}

\subsection{Research Question 3: Models without case estimate data versus case estimates} \label{q3}

While the previous question was concerned with the value of case estimates as an input to a model, our final research question instead asks whether a model trained without case estimate data is comparable to the case estimates themselves. Here we discuss the performance of the FNN against the case estimates. When considering case estimates as a model for this research question, we are referring to the latest simulated estimate of the outstanding cost for each claim at the valuation date.

\begin{table}[!htb]
    \centering
    \begin{tabular}{@{}crrrrc@{}}
    \toprule
                   & $\lvert \text{ResErr} \rvert$ & MALE  & MSLE \\ \midrule
    Test statistic & 10                            & 34    & 50    \\
    $p$-value      & $7.6 \times 10^{-14}$ & $6.6 \times 10^{-12}$ & $5.6 \times 10^{-11}$  \\ 
    Win count      & 47                            & 45    & 49   \\\bottomrule
    \end{tabular}
    \caption{Wilcoxon two-sided test statistic and $p$-value comparing the FNN and case estimates for each evaluation metric. The win count represents the number of datasets (out of 50) where the FNN produced a smaller value than the case estimates for the absolute reserve error, MALE and MSLE metrics.}
    \label{wilcoxon q3}
\end{table}

Compare results of `Case Estimates' with that of `FNN` in Table \ref{results table q2}. 
On the one hand, the FNN leads to a smaller mean reserve error and smaller mean MSLE which suggest the FNN outperforms the case estimates at both an aggregate and individual level. However, the MALE is similar between the two and the $\text{vsCE}_\text{O}$ is hovering around 50\% (refer to Table \ref{vsCE table q2}), which suggests ranking both approaches is more difficult at an individual claim level. However, when considering the results in Table \ref{wilcoxon q3}, it becomes clearer that on any individual test dataset, the FNN tends to outperform the case estimates.

\begin{figure}[!htb]
    \centering
    \subfloat[FNN]{
        \includegraphics[height=4.5cm, keepaspectratio]{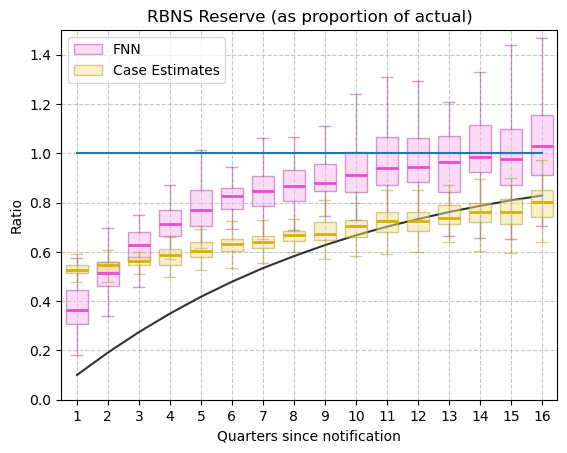}
        \label{fig:FNN_CE_OCLs_dev}
    }
    \subfloat[FNN]{
        \includegraphics[height=4.5cm, keepaspectratio]{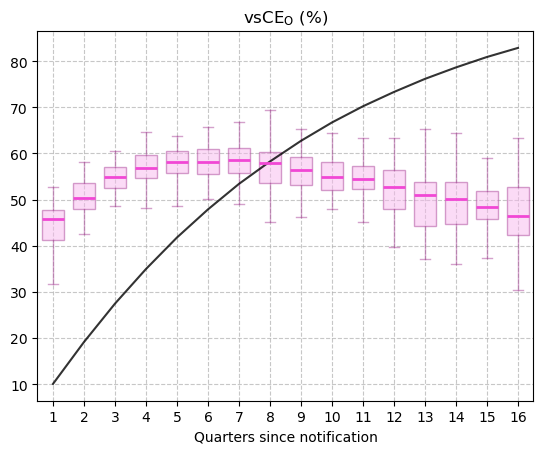}
        \label{fig:FNN_CE_vsCE_OCL_dev}
    }
    \caption{Boxplots showing the aggregate predictions of outstanding claim amounts (expressed as a proportion of the true amounts outstanding) and $\text{vsCE}_\text{O}$ for the FNN and case estimates at the valuation date, subset by both accident quarter and quarters since notification. The black curves represent the proportion of the actual amounts outstanding related to predictions made on or before quarter q. Stronger performance is indicated by values closer to 1 for \protect\subref{fig:FNN_CE_OCLs_dev} and larger values for \protect\subref{fig:FNN_CE_vsCE_OCL_dev}.}
    \label{fig: FNN}
\end{figure}

\begin{figure}[!htb]
    \centering
    \subfloat[MALE]{
        \includegraphics[height=4.5cm, keepaspectratio]{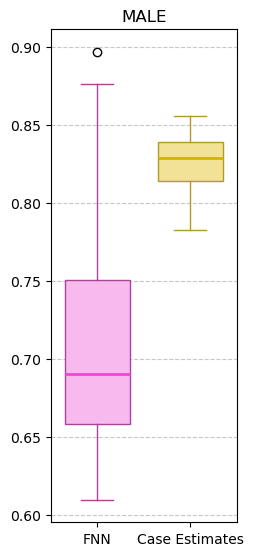}
        \label{fig:FNN_MALE_val}
    }
    \subfloat[MSLE]{
        \includegraphics[height=4.5cm, keepaspectratio]{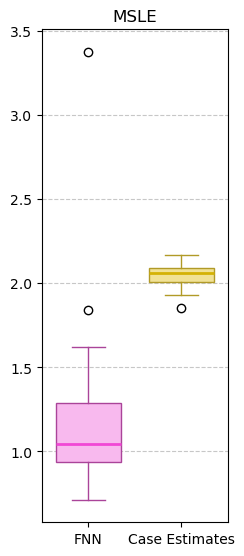}
        \label{fig:FNN_MSLE_val}
    }
    \subfloat[ResErr]{
        \includegraphics[height=4.5cm, keepaspectratio]{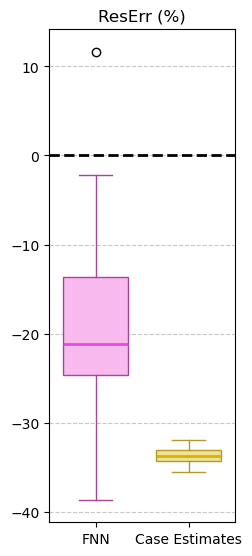}
        \label{fig:FNN_OCLs_val}
    }
    \subfloat[$\text{vsCE}_\text{O}$]{
        \includegraphics[height=4.5cm, keepaspectratio]{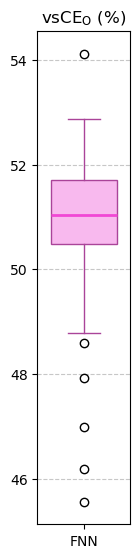}
        \label{fig:FNN_vsCE_OCL_val}
    }

    \caption{Boxplots containing MALE, MSLE, reserve error and $\text{vsCE}_\text{O}$ metrics for the FNN and case estimates at the valuation date. Smaller values for \protect\subref{fig:FNN_MALE_val} and \protect\subref{fig:FNN_MSLE_val}, values closer to 0 for \protect\subref{fig:FNN_OCLs_val} and larger values for \protect\subref{fig:FNN_vsCE_OCL_val} indicate stronger performance.}
    \label{fig:boxplots-val FNN}
\end{figure}

\noindent Figure \ref{fig: FNN} provides some more insight into the differences between these two `models'. For the first few quarters of a claim's development, the FNN performs particularly poorly. This is because in these first few quarters, there would be little available payment information and without case estimate data, the FNN does not have enough information to make a reasonable forecast. In practice, however, it is likely that there would be many static covariates that could be input to the model to assist particularly in these early development periods.

As mentioned earlier, the FNN compensates for the early underestimation by overestimating in later development quarters. Ultimately, the plots suggest that the FNN outperforms the case estimates from around 5-15 quarters after notification, which accounts for around 40\% of the true amounts outstanding at the valuation date. 

Figure \ref{fig:boxplots-val FNN} contains boxplots of the MALE, MSLE, reserve error and $\text{vsCE}_\text{O}$ metrics at the valuation date. These graphs suggest that the predictions produced by the FNN are more volatile than those produced by the case estimates. Regardless, the MALE, MSLE and reserve error all tend to be smaller for the FNN more often than not.

The answer to this question, then, is somewhat mixed. The results suggest that the performance between the two `models' is comparable in this scenario where case estimates, although materially downward-biased in aggregate, remain informative at an individual claim level.

\subsection{Sensitivity analysis with distorted case estimate data} \label{distorted}

Although the primary contribution of this paper lies in the framework used to answer the stated research questions, we present a small sensitivity analysis to assess the robustness of the models in environments where case estimate data is of poorer quality. 

This experiment is performed on the same complexity 5 datasets presented in Sections \ref{q1} to \ref{q3} under the same train-test splits. The only difference is that for each transaction in each observation, an adjustment of $(1 - k_c \cdot u_t)$ is applied to the case estimate at the time of the transaction, where $k_c \sim U(0.7, 0.9)$ is the adjustment factor per claim $c$ and $u_t$ is the proportion of the claim’s lifetime still outstanding at transaction time $t$. $u_t$ uses the realised (ex post) lifetime to determine the proportion outstanding. This way, case estimates will continue to be strictly positive while a claim is open. As $u_t \rightarrow 0$, the adjustment factor approaches 1, meaning the distortion effectively vanishes. At notification, case estimates will on average be significantly understated. It also means that incurred amount, that is, the payments to date plus the latest case estimate, will increase on average as a claim develops. Since this adjustment applies at all transaction times, including where there is no existing case estimate revision, this will effectively increase the number of transactions with case estimate revisions. This approach was taken to mitigate the risk of a negative case estimate. Similarly, the adjustment was applied to the case estimates directly instead of the total incurred amount to prevent the emergence of negative case estimates. The adjustment factor $k_c$ randomly varies for each claim to add further noise to the case estimate data.

The FNN+ from Section \ref{q1} undergoes the same hyperparameter tuning and training processes with the distorted case estimate data. The selected hyperparameter combination can be found in the supplementary material.

\begin{table}[!htb]
    \centering
    \begin{tabular}{@{}crrrrc@{}}
    \toprule
    Model     & ResErr (\%) & MALE          & MSLE  \\ \midrule
    Case Estimates & -33.6 (0.8)    & 0.828 (0.017) & 2.051 (0.066)   \\
    Case Estimates Distorted & -63.5 (0.6)    & 1.490 (0.023) & 4.538 (0.111) \\
    FNN+      & -7.7 (6.9)     & 0.550 (0.036) & 0.633 (0.068)   \\
    FNN+ Distorted & -13.4 (6.1) & 0.580 (0.044) & 0.680 (0.096)   \\ 
    FNN       & -19.3 (9.2)    & 0.712 (0.070) & 1.143 (0.400)   \\ \bottomrule
    \end{tabular}
    \caption{Reserve error, MALE and MSLE metrics for the regular and distorted case estimates, the FNN+ with regular and distorted case estimate data, and the FNN at the valuation date across 50 datasets with distorted case estimate data. Values are expressed as mean (standard deviation).}
    \label{results table distorted}
\end{table}

\begin{table}[!htb]
    \centering
    \begin{tabular}{@{}crrrrc@{}}
    \toprule
                   & $\lvert \text{ResErr} \rvert$ & MALE  & MSLE \\ \midrule
    Test statistic & 181                           & 328   & 374  \\
    $p$-value      & $2.8 \times 10^{-6}$ & 0.0024 & 0.0103 \\ 
    Win count      & 37                            & 33    & 32   \\ \bottomrule
    \end{tabular}
    \caption{Wilcoxon two-sided test statistic and $p$-value comparing the FNN+ with regular and distorted case estimate data for each evaluation metric. The win count represents the number of datasets (out of 50) where the FNN+ with regular case estimate data produced a smaller value than the FNN+ with distorted case estimate data for the absolute reserve error, MALE and MSLE metrics.}
    \label{wilcoxon distorted 1}
\end{table}

\begin{table}[!htb]
    \centering
    \begin{tabular}{@{}crrrrc@{}}
    \toprule
                   & $\lvert \text{ResErr} \rvert$ & MALE  & MSLE \\ \midrule
    Test statistic & 246                           & 9     & 2    \\
    $p$-value      & $8.6 \times 10^{-5}$ & $5.9 \times 10^{-14}$ & $5.3 \times 10^{-15}$ \\
    Win count      & 37                             & 48    & 49   \\ \bottomrule
    \end{tabular}
    \caption{Wilcoxon two-sided test statistic and $p$-value comparing the FNN+ with distorted case estimate data and the FNN for each evaluation metric. The win count represents the number of datasets (out of 50) where the FNN+ with distorted case estimate data produced a smaller value than the FNN for the absolute reserve error, MALE and MSLE metrics.}
    \label{wilcoxon distorted 2}
\end{table}

Consider Table \ref{results table distorted}. While the regular case estimates already display a mean downward bias of 33.6\%, this increases to 63.5\% after the transformation is applied. Similarly, the mean MALE almost doubles and the mean MSLE more than doubles after distortion. It is clear that the quality of the case estimate data has deteriorated significantly from the adjustment. 

When examining the FNN+ with distorted case estimate data, however, it still significantly outperforms the FNN with no case estimate data, as evidenced by its smaller mean reserve error, MALE and MSLE. Compared to the FNN+ with regular case estimate data, there is a slight deterioration across all metrics, yet it is much less pronounced than that observed between the regular and distorted case estimates. Tables \ref{wilcoxon distorted 1} and \ref{wilcoxon distorted 2} all contain similar conclusions. Even with this distorted case estimate data, the FNN+ continues to significantly outperform the FNN on the absolute reserve error, MALE and MSLE metrics. When comparing the FNN+ under both regular and distorted case estimate data, all metrics suggest the distortion deteriorates the performance of the FNN+ at a 5\% level of significance.

\begin{figure}[!htb]
    \centering
    \subfloat[]{
        \includegraphics[height=4.4cm, keepaspectratio]{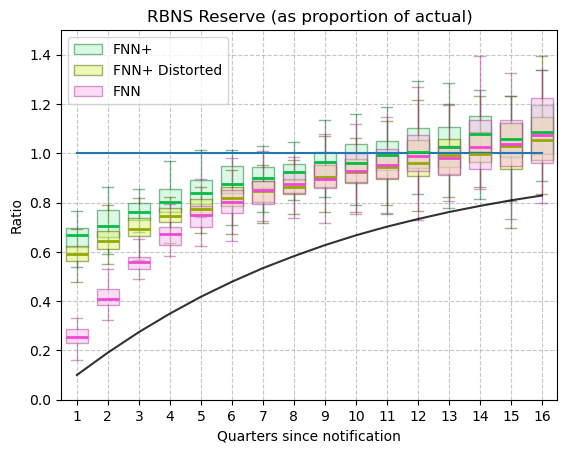}
        \label{fig:FNNplus_FNNplus_Distorted_FNN_OCLs_dev}
    }
    \subfloat[]{
        \includegraphics[height=4.4cm, keepaspectratio]{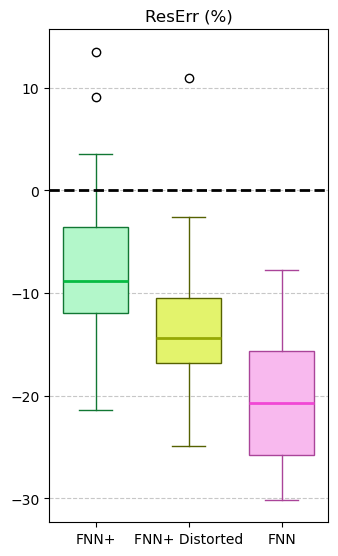}
        \label{fig:FNNplus_FNNplus_Distorted_FNN_OCLs_val}
    }
    \subfloat[]{
        \includegraphics[height=4.4cm, keepaspectratio]{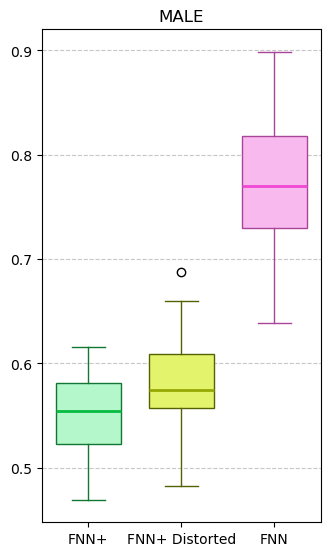}
        \label{fig:FNNplus_FNNplus_Distorted_FNN_MALE_val}
    }
    \subfloat[]{
        \includegraphics[height=4.4cm, keepaspectratio]{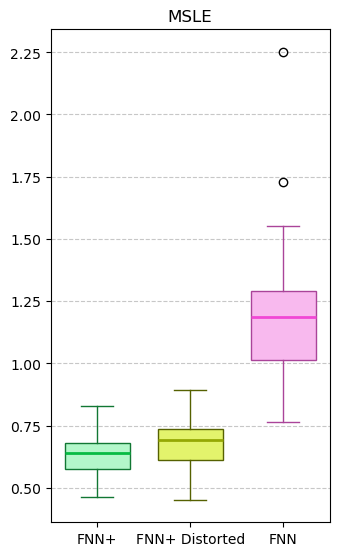}
        \label{fig:FNNplus_FNNplus_Distorted_FNN_MSLE_val}
    }
    \caption{Boxplots showing the aggregate predictions of outstanding claim amounts (expressed as a proportion of the true amounts outstanding) for the FNN+ with regular and distorted case estimate data and the FNN at the valuation date, subset by quarters since notification, as well as the reserve error, MALE and MSLE for these models at the valuation date. The black curve represents the proportion of the actual amounts outstanding related to predictions made on or before quarter q. Stronger performance is indicated by values closer to 1 for \protect\subref{fig:FNNplus_FNNplus_Distorted_FNN_OCLs_dev} and values closer to 0 for \protect\subref{fig:FNNplus_FNNplus_Distorted_FNN_OCLs_val}, \protect\subref{fig:FNNplus_FNNplus_Distorted_FNN_MALE_val} and \protect\subref{fig:FNNplus_FNNplus_Distorted_FNN_MSLE_val}.}
    \label{fig:FNNplus_FNNplus_Distorted_FNN}
\end{figure} 

Figure \ref{fig:FNNplus_FNNplus_Distorted_FNN} explores these trends in further detail. It becomes apparent that the main differences in the reserve errors are heavily concentrated in the first 4 quarters since notification. This aligns with expectations as the downward adjustment to the case estimates is strongest during early stages of a claim's development, and so the distortions to the FNN+ are likely to be most impactful in this maturity range. However, in this environment the distorted case estimates still appear to add value to the FNN+ relative to the FNN, particularly for immature claims.

When looking at Figure \ref{fig:FNNplus_FNNplus_Distorted_FNN_OCLs_val}, the reserve error of the distorted FNN+ is distributed rather evenly between the distributions of the two other models. For the MALE and MSLE plots, however, the distorted FNN+ is much more aligned to the regular FNN+ as opposed to the FNN. This suggests that this particular downward distortion of case estimate data has negatively affected the predictions of the FNN+ at an aggregate level more so than at an individual claim level.

Ultimately, since the FNN+ relies on input case estimate data, it is contaminated by case estimate inaccuracy, but it attenuates the inaccuracy reasonably well in this environment. 

\subsection{Sensitivity analysis with noisy case estimate data} \label{noisy}

Similar to Section \ref{distorted}, we have included another sensitivity analysis to assess the impact of poorer quality case estimate data. Again, this experiment is performed on the same complexity 5 datasets presented in Sections \ref{q1} to \ref{q3} under the same train-test splits. 

Instead of applying a downward bias to the case estimate data, as done in Section \ref{distorted}, random noise is added to the case estimate data, which can be positive or negative. More formally, we define the case estimate development process, $O_c(t)$ as a continuous-time stochastic process that represents the case estimate of claim $c$ at time $t$ since claim notification. We introduce noise into this process as

\begin{equation}
    O^{noisy}_c(t) = a_{c,t} \cdot O_c(t),
\end{equation} where $a_{c,t} \sim U(0.7, 1.3)$ is a piecewise-constant stochastic process that is independently resampled at each major or minor case estimate revision as defined in SPLICE \citep*{AvTaWaWo21}, and remains unchanged until the next revision. This does, however, open the possibility for negative case estimates to emerge if $a_{c,t_1}<1$ which is followed immediately by a payment at $t_2$ being greater than $a_{c,t_2} \cdot O_c(t_2) = a_{c,t_1} \cdot O_c(t_2), \; t_2 = t_1 + \epsilon$. In such cases, we adjust the case estimate upward so that $O^{noisy}_c(t_2) = 0.1 \cdot O^{noisy}_c(t_1)$ and remains as such until the next revision or payment occurs.

The hyperparameter tuning and model training processes for the FNN+ with noisy case estimate data are the same as those used for the other models. The selected hyperparameter combination can be found in the supplementary material.

\begin{table}[!htb]
    \centering
    \begin{tabular}{@{}crrrrc@{}}
    \toprule
    Model     & ResErr (\%) & MALE          & MSLE  \\ \midrule
    Case Estimates & -33.6 (0.8)    & 0.828 (0.017) & 2.051 (0.066)   \\
    Case Estimates Noisy & -33.4 (0.9)    & 0.908 (0.017) & 2.134 (0.069) \\
    FNN+      & -7.7 (6.9)     & 0.550 (0.036) & 0.633 (0.068)   \\
    FNN+ Noisy & -4.3 (36.3) & 0.564 (0.043) & 0.682 (0.175)   \\ 
    FNN       & -19.3 (9.2)    & 0.712 (0.070) & 1.143 (0.400)   \\ \bottomrule
    \end{tabular}
    \caption{Reserve error, MALE and MSLE metrics for the regular and noisy case estimates, the FNN+ with regular and noisy case estimate data, and the FNN at the valuation date across 50 datasets with noisy case estimate data. Values are expressed as mean (standard deviation).}
    \label{results table noisy}
\end{table}

\begin{table}[!htb]
    \centering
    \begin{tabular}{@{}crrrrc@{}}
    \toprule
                   & $\lvert \text{ResErr} \rvert$ & MALE  & MSLE \\ \midrule
    Test statistic & 436                           & 524   & 543  \\
    $p$-value      & 0.0518 & 0.2780 & 0.3673 \\ 
    Win count      & 33                            & 27    & 25   \\ \bottomrule
    \end{tabular}
    \caption{Wilcoxon two-sided test statistic and $p$-value comparing the FNN+ with regular and noisy case estimate data for each evaluation metric. The win count represents the number of datasets (out of 50) where the FNN+ with regular case estimate data produced a smaller value than the FNN+ with noisy case estimate data for the absolute reserve error, MALE and MSLE metrics.}
    \label{wilcoxon noisy 1}
\end{table}

\begin{table}[!htb]
    \centering
    \begin{tabular}{@{}crrrrc@{}}
    \toprule
                   & $\lvert \text{ResErr} \rvert$ & MALE  & MSLE \\ \midrule
    Test statistic & 193                           & 5     & 19   \\
    $p$-value      & $5.7 \times 10^{-6}$ & $1.8 \times 10^{-14}$ & $5.5 \times 10^{-13}$ \\
    Win count      & 40                             & 48    & 49   \\ \bottomrule
    \end{tabular}
    \caption{Wilcoxon two-sided test statistic and $p$-value comparing the FNN+ with noisy case estimate data and the FNN for each evaluation metric. The win count represents the number of datasets (out of 50) where the FNN+ with noisy case estimate data produced a smaller value than the FNN for the absolute reserve error, MALE and MSLE metrics.}
    \label{wilcoxon noisy 2}
\end{table}

Compare Table \ref{results table distorted} to Table \ref{results table noisy}. While the previous distortion contributed to a mean downward bias of 63.5\%, the noise adjustment in this section hardly alters the mean reserve error in comparison. Similarly, while the mean MALE and mean MSLE do slightly increase with the additional noise, they do not increase to the same extent as that observed under the previous distortion. From this angle, the quality of the case estimate data has not deteriorated to the same extent as it had from the previous distortion. 

As expected, the FNN+ with noisy case estimate data still significantly outperforms the FNN with no case estimate data, as evidenced by its smaller mean reserve error, MALE and MSLE. Compared to the FNN+ with regular case estimate data, there is a slight deterioration across the MALE and MSLE, while the mean reserve error instead improves with the noise adjustment. Upon further inspection, this is primarily due to a single outlier that generated a reserve error of almost 250\%. This outlier is also contributing to the relatively high reserve error variance.

While Table \ref{wilcoxon noisy 2} reinforces that the FNN+ with noisy case estimate data continues to significantly outperform the FNN, Table \ref{wilcoxon noisy 1} instead highlights some interesting observations. The 30\% noise adjustment is not enough to significantly deteriorate the mean performance of the FNN+ across any of the metrics presented in the table.

\begin{figure}[!htb]
    \centering
    \subfloat[]{
        \includegraphics[height=4.4cm, keepaspectratio]{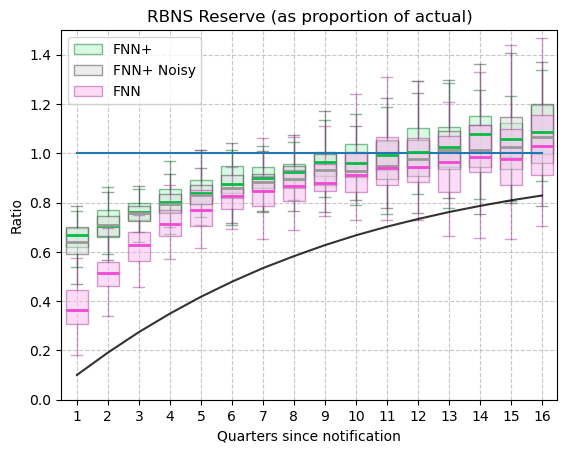}
        \label{fig:FNNplus_FNNplus_Noisy_FNN_OCLs_dev}
    }
    \subfloat[]{
        \includegraphics[height=4.4cm, keepaspectratio]{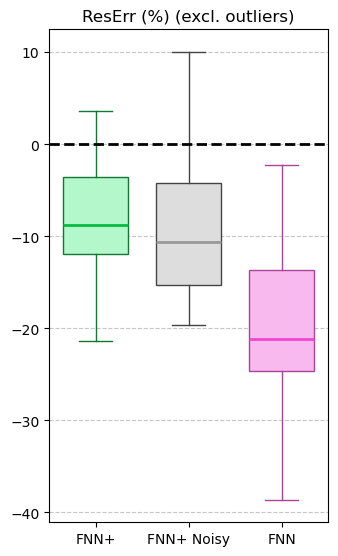}
        \label{fig:FNNplus_FNNplus_Noisy_FNN_OCLs_val}
    }
    \subfloat[]{
        \includegraphics[height=4.4cm, keepaspectratio]{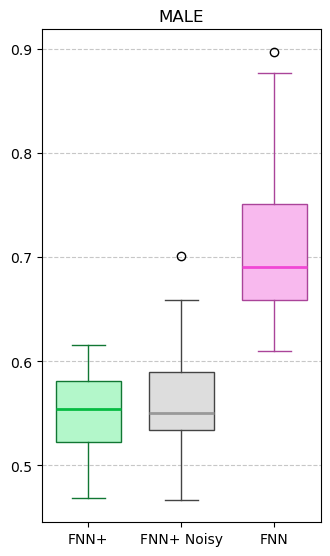}
        \label{fig:FNNplus_FNNplus_Noisy_FNN_MALE_val}
    }
    \subfloat[]{
        \includegraphics[height=4.4cm, keepaspectratio]{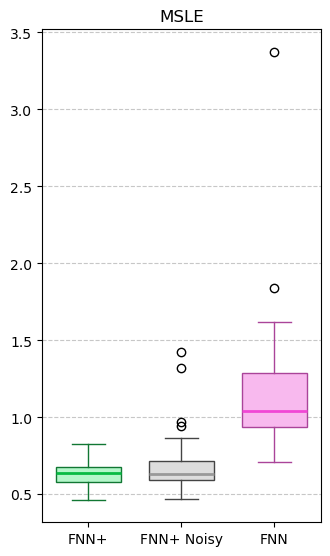}
        \label{fig:FNNplus_FNNplus_Noisy_FNN_MSLE_val}
    }
    \caption{Boxplots showing the aggregate predictions of outstanding claim amounts (expressed as a proportion of the true amounts outstanding) for the FNN+ with regular and noisy case estimate data and the FNN at the valuation date, subset by quarters since notification, as well as the reserve error, MALE and MSLE for these models at the valuation date. The black curve represents the proportion of the actual amounts outstanding related to predictions made on or before quarter q. Stronger performance is indicated by values closer to 1 for \protect\subref{fig:FNNplus_FNNplus_Noisy_FNN_OCLs_dev} and values closer to 0 for \protect\subref{fig:FNNplus_FNNplus_Noisy_FNN_OCLs_val}, \protect\subref{fig:FNNplus_FNNplus_Noisy_FNN_MALE_val} and \protect\subref{fig:FNNplus_FNNplus_Noisy_FNN_MSLE_val}.}
    \label{fig:FNNplus_FNNplus_Noisy_FNN}
\end{figure} 

Figure \ref{fig:FNNplus_FNNplus_Noisy_FNN_OCLs_dev} examines differences in forecasted reserve predictions by quarters since claim notification. Since the noise adjustment does not directly scale in absolute terms with the maturity of each claim at each of its revision times, it is expected that any deviations it creates in the reserve forecasts of the FNN+ would also be evenly distributed by quarter since claim notification. Indeed, this is what the results suggest, as there does not seem to be any persistent over- or under-estimation relative to the unaltered FNN+ by quarter since claim notification. Across all plots in Figure \ref{fig:FNNplus_FNNplus_Noisy_FNN}, the FNN+ with noisy case estimates is more aligned to the FNN+ with regular case estimate data as opposed to the FNN. The aforementioned outlier has been removed from Figure \ref{fig:FNNplus_FNNplus_Noisy_FNN_OCLs_val}, however. Although the central estimates are closely aligned for the FNN+ trained under both regular and noisy case estimate data, the latter tends to display a higher variance across all metrics. Again, this is somewhat expected as the noise adjustment is not biased upward or downward.

Ultimately, the FNN+ has been contaminated by the additional noise in the case estimate data but to a lesser degree than the previous distortion. All versions of case estimate data presented in this paper possessed strong downward biases in aggregate, which may not be the case for all insurers. Some may have more conservative, noisier, or otherwise different patterns embedded in their case estimate data and should be a strong motivator for insurers to repeat this experiment using their own data.

\section{Conclusion} \label{Conclusion}

In this paper, we produced a framework for answering three main research questions that have been overlooked in the literature thus far. Our questions, and their answers in our simulated environment, are as follows:
\begin{enumerate}
    \item Is there a significant difference in performance between a feed-forward network trained solely on summary data compared to an LSTM trained on full transactional data? 
    
    \textbf{The transactional data appears to provide a slight benefit to the reserve forecasts, but does not improve the results at an individual claim level.}
    \item Does a neural network (recurrent or feed-forward) trained with case estimate data perform better than the same model without case estimates? 
    
    \textbf{Yes, resoundingly. In the absence of additional covariates, this result appears to hold even if the case estimate data is of poorer quality.}
    \item Does a neural network (recurrent or feed-forward) trained without case estimate data outperform the case estimates? 
    
    \textbf{The FNN is broadly competitive with, and in several respects improves upon, raw case estimates, depending on claim maturity.}
\end{enumerate} 
Although we believe our methodology is robust, we acknowledge the following limitations:
\begin{enumerate}
    \item When using the data generated by SPLICE \citep*{AvTaWaWo21,AvTaWa23}, all that is available for each claim is the (calendar and development) times of payments, case estimate histories and three basic covariates. In practice, insurers would possess more detailed information about each claim, such as the name of the claim assessor, written claims descriptions, geographical location and other line-of-business-specific data that could potentially be useful for estimating future claims costs.

    \item Due to the nature of individual loss reserving, only reported claims have been considered. Therefore, IBNR claims will need to be modelled separately.

    \item The models only produce point estimates and hence are devoid of the benefits of stochastic models.

    \item While models have been compared both with and without case estimate data in Research Question 2, this investigation of input feature importance could be expanded through the use of explainable AI tools such as SHAP \citep*{LuLe17}. We do acknowledge, however, that SHAP is a computationally expensive tool and is not trivial to implement in a time-series environment as the standard implementations rest on feature-independence assumptions that sit awkwardly with autocorrelated transactional inputs.

    \item Although the results of the models in isolation are not the primary focus of this paper, using more complex variants of feed-forward or recurrent network structures, such as a Distributional Refinement Network \citep*{AvDoLaWo25} or Transformer \citep{VaShPaUsJoGoKaPo17}, respectively, could lead to different conclusions.
    
    \item Additionally, further hyperparameter tuning could be investigated if one wanted to maximise the performance of the models for practical use. This could be done by testing a wider range of values for each hyperparameter, employing Fixed Origin or Rolling Origin Cross-Validation \citep{Tas00} instead of allocating resources to a standalone validation set, and perhaps by adjusting the criteria used to select the `best' hyperparameter combination for each model.

    \item Runtime costs are not considered within an explicit evaluation metric but they should have some influence on the relative merits of each model in practice.
\end{enumerate} 
Ultimately, we must stress that the results presented in this paper and our answers to the research questions may not reflect any insurer's actual circumstances. Therefore, we would like to emphasise our framework and our approach to answering these questions as the key contributions of the paper and encourage others to perform these tests in their own environments.

\section*{Acknowledgements}

Avanzi and Wong acknowledge support under Australian Research Council's Discovery Project \linebreak (DP200101859) funding scheme. The views expressed herein are those of the authors and are not necessarily those of the supporting organisations. The authors declare no conflicts of interest regarding this article.

\section*{Data and Code}

All results in this paper are reproducible using Python and R, with code and data available on Zenodo (\url{https://zenodo.org/records/18005906}) and GitHub (\url{https://github.com/agi-lab/reserving-RNN}). Datasets were simulated in R, while modelling was conducted using Python. Hyperparameter tuning and model training was run in an Azure ML notebook on a Standard E8s v3 virtual machine with 8 cores, while the dataset simulation and evaluation of models on the test set were run on a local machine with an Nvidia RTX 3070 GPU and an AMD Ryzen 5 5600X CPU with 6 cores. The Jupyter Notebooks with the results of the tuned models on the validation and test sets are available on both Zenodo and Github. Creation and processing of datasets took roughly 40 hours. It took approximately 450 hours to tune and train all models. Evaluation of all models, including those in the appendix and supplementary material, on the test sets required around 40 hours.

\section*{References}

\bibliographystyle{elsarticle-harv}
\bibliography{libraries}

\appendix

\section{Selection of validation set results} \label{results validation}

\begin{figure}[H]
    \centering
    \subfloat[LSTM+]{
        \includegraphics[width=0.275\textwidth]{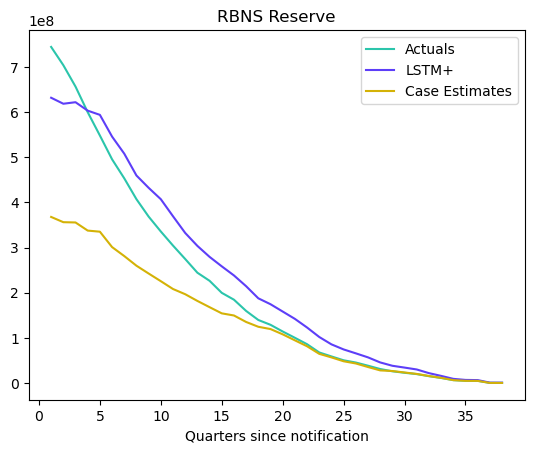}
        \label{fig:LSTMplus_OCLs_dev_validation}
    }
    \subfloat[FNN+]{
        \includegraphics[width=0.275\textwidth]{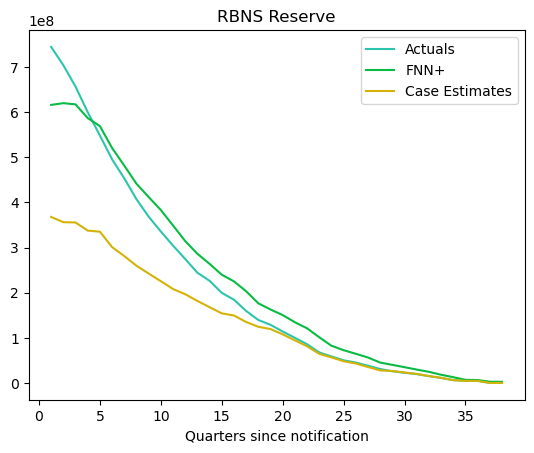}
        \label{fig:FNNplus_OCLs_dev_validation}
    }
    \\
    \subfloat[LSTM]{
        \includegraphics[width=0.275\textwidth]{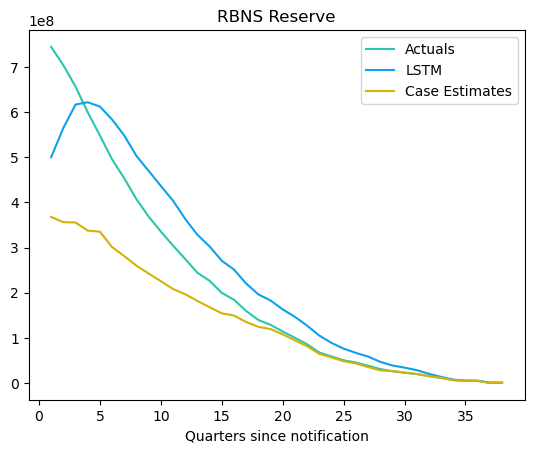}
        \label{fig:LSTM_OCLs_dev_validation}
    }
    \subfloat[FNN]{
        \includegraphics[width=0.275\textwidth]{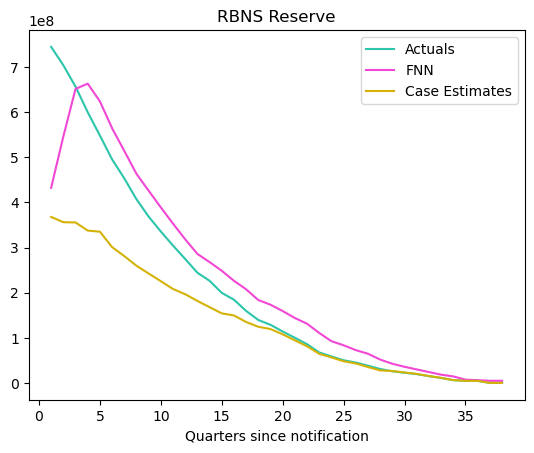}
        \label{fig:FNN_OCLs_dev_validation}
    }
    
    \caption{Curves showing the aggregate predictions of outstanding claim amounts from the tuned models in the validation set, subset by quarters since notification.}
    \label{fig:OCLs_dev_validation}

\end{figure}

\begin{figure}[H]

    \centering
    \subfloat[LSTM+]{
        \includegraphics[width=0.275\textwidth]{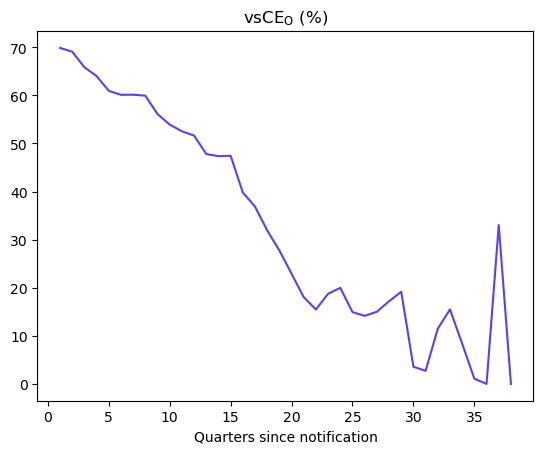}
        \label{fig:LSTMplus_vsCE_OCL_dev_validation}
    }
    \subfloat[FNN+]{
        \includegraphics[width=0.275\textwidth]{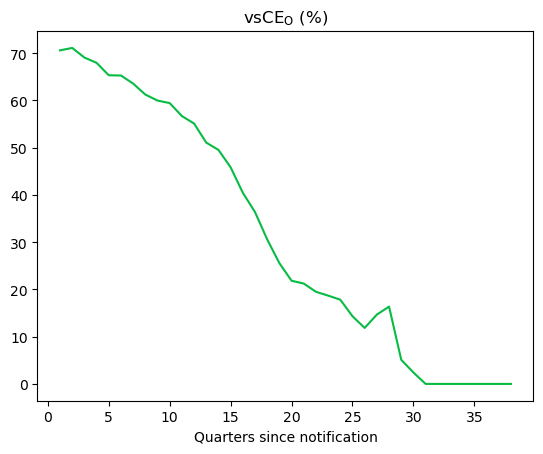}
        \label{fig:FNNplus_vsCE_OCL_dev_validation}
    }
    \\
    \subfloat[LSTM]{
        \includegraphics[width=0.275\textwidth]{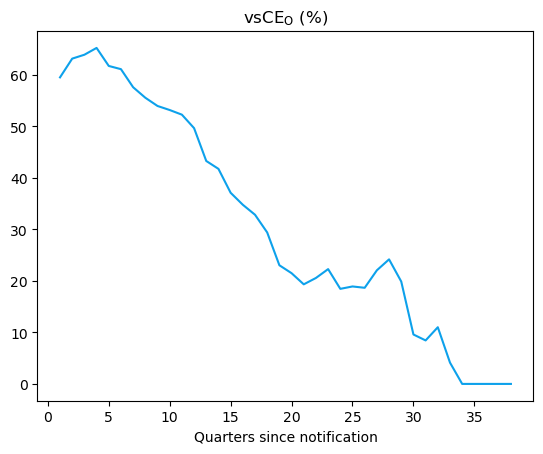}
        \label{fig:LSTM_vsCE_OCL_dev_validation}
    }
    \subfloat[FNN]{
        \includegraphics[width=0.275\textwidth]{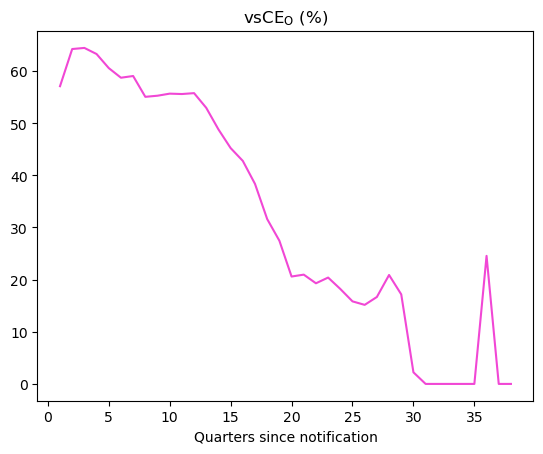}
        \label{fig:FNN_vsCE_OCL_dev_validation}
    }
    
    \caption{Curves showing the $\text{vsCE}_\text{O}$ from the tuned models in the validation set, subset by quarters since notification.}
    \label{fig:vsCE_OCL_dev_validation}
\end{figure}

This section presents a selection of results from each model using all available observations in the validation set that was used for hyperparameter tuning. The values in Table \ref{results table validation} and Figures \ref{fig:OCLs_dev_validation} and \ref{fig:vsCE_OCL_dev_validation} are from a single training run of each model using its selected hyperparameter combination.

\begin{table}[!htb]
    \centering
    \begin{tabular}{@{}crrrrc@{}}
    \toprule
    Model          & ResErr (\%) & MALE  & MSLE  & $\text{vsCE}_\text{O}$ (\%) \\ \midrule
    Case Estimates & -34.3          & 0.694 & 1.358 & -         \\
    LSTM+          & 11.7           & 0.331 & 0.187 & 54.7      \\ 
    FNN+           & 7.6            & 0.313 & 0.182 & 57.1      \\ 
    LSTM           & 14.4           & 0.426 & 0.309 & 51.9      \\
    FNN            & 9.6            & 0.427 & 0.337 & 53.0      \\ \bottomrule
    \end{tabular}
    \caption{Reserve error, MALE, MSLE and $\text{vsCE}_\text{O}$ metrics for the tuned models in the validation set.}
    \label{results table validation}
\end{table}

\section{Additional results by quarter since notification}

\begin{figure}[H]
    \centering
    \subfloat[LSTM+]{
        \includegraphics[width=0.275\textwidth]{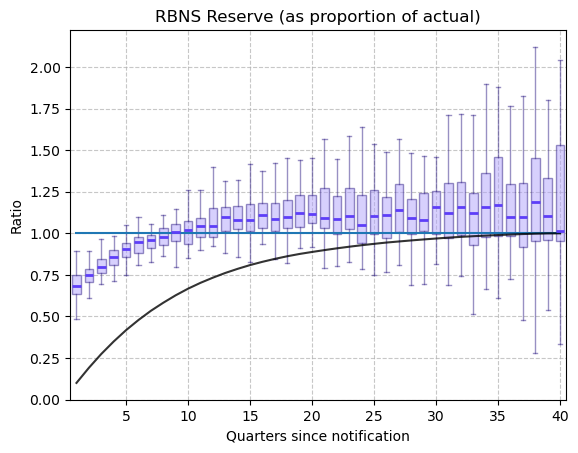}
        \label{fig:LSTMplus_OCLs_dev}
    }
    \subfloat[FNN+]{
        \includegraphics[width=0.275\textwidth]{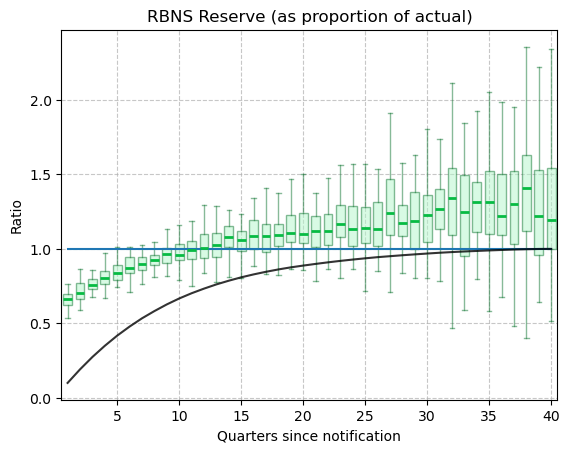}
        \label{fig:FNNplus_OCLs_dev}
    }
    \subfloat[LSTM]{
        \includegraphics[width=0.275\textwidth]{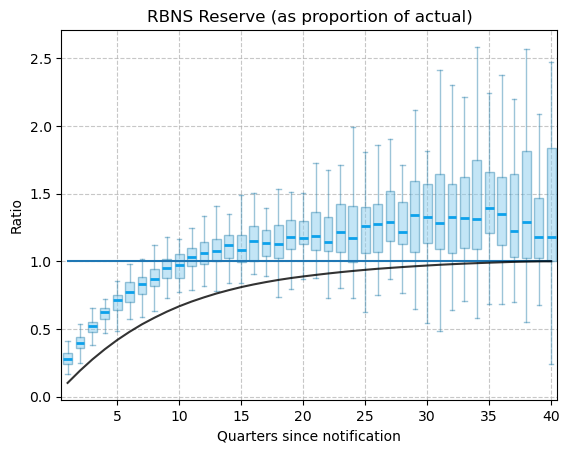}
        \label{fig:LSTM_OCLs_dev}
    }
    \\
    \subfloat[FNN]{
        \includegraphics[width=0.275\textwidth]{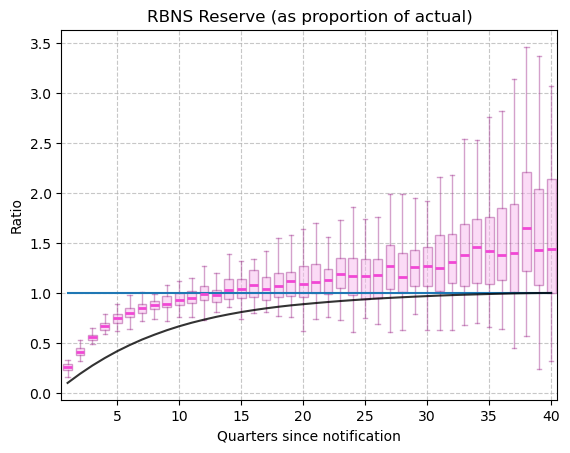}
        \label{fig:FNN_OCLs_dev}
    }
    \subfloat[Case Estimates]{
        \includegraphics[width=0.275\textwidth]{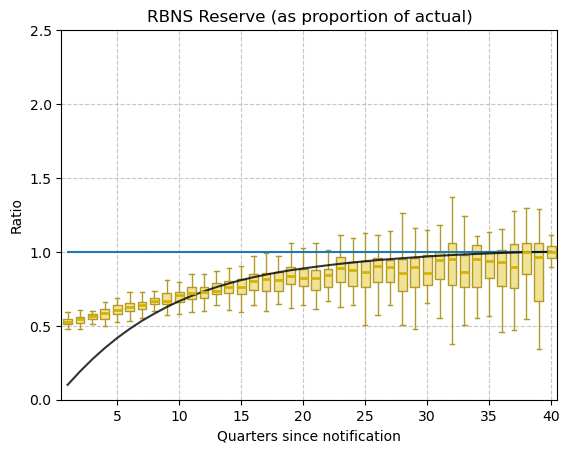}
        \label{fig:CE_OCLs_dev}
    }
    
    \caption{Boxplots showing the aggregate predictions of outstanding claim amounts (expressed as a proportion of the true amounts outstanding) at the valuation date, subset by quarters since notification. The black curves represent the proportion of the actual amounts outstanding related to predictions made on or before quarter q.}
    \label{fig:OCLs_dev}

\end{figure}

\begin{figure}[H]
    \centering
    \subfloat[LSTM+]{
        \includegraphics[width=0.275\textwidth]{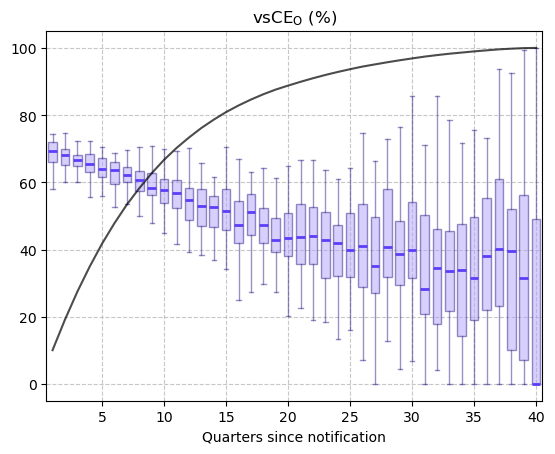}
        \label{fig:LSTMplus_vsCE_OCL_dev}
    }
    \subfloat[FNN+]{
        \includegraphics[width=0.275\textwidth]{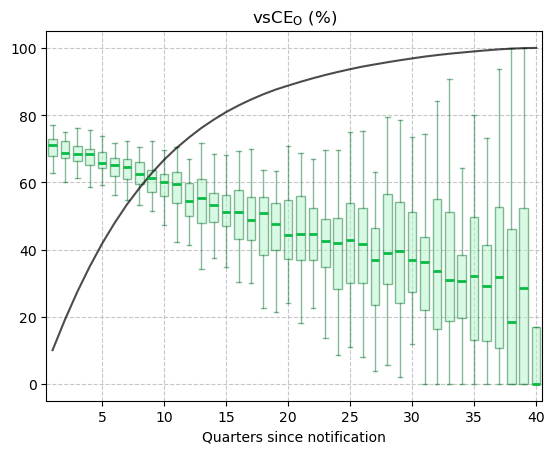}
        \label{fig:FNNplus_vsCE_OCL_dev}
    }
    \\
    \subfloat[LSTM]{
        \includegraphics[width=0.275\textwidth]{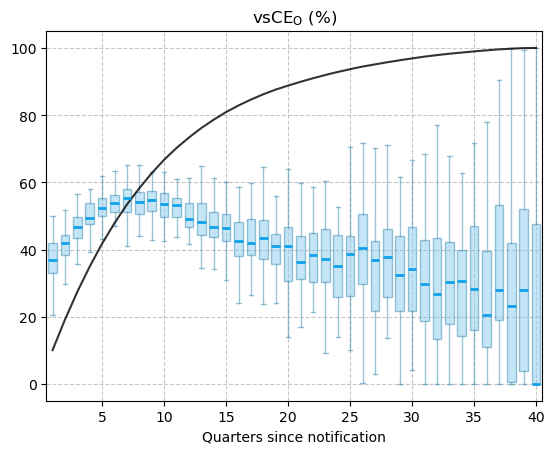}
        \label{fig:LSTM_vsCE_OCL_dev}
    }
    \subfloat[FNN]{
        \includegraphics[width=0.275\textwidth]{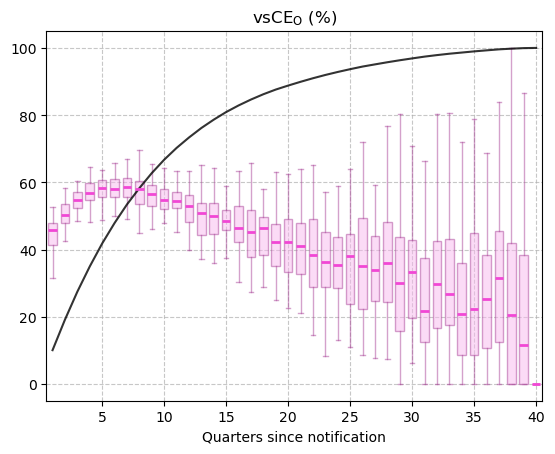}
        \label{fig:FNN_vsCE_OCL_dev}
    }
    
    \caption{Boxplots showing the $\text{vsCE}_\text{O}$ at the valuation date, subset by quarters since notification. The black curves represent the proportion of the actual amounts outstanding related to predictions made on or before quarter q.}
    \label{fig:vsCE_OCL_dev}
\end{figure}

\section{Additional results by accident quarter}

\begin{figure}[H]
    \centering
    \subfloat[LSTM+]{
        \includegraphics[width=0.275\textwidth]{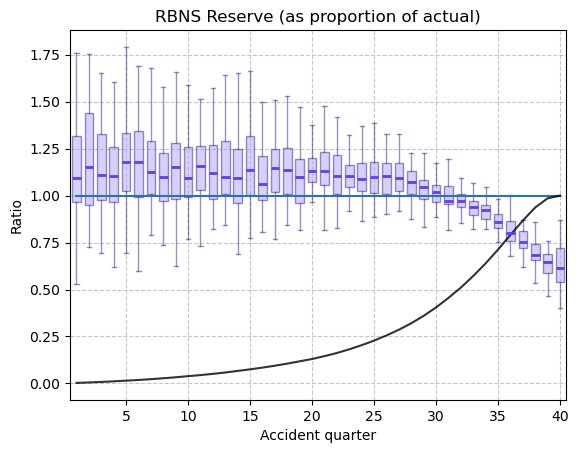}
        \label{fig:LSTMplus_OCLs_acc}
    }
    \subfloat[FNN+]{
        \includegraphics[width=0.275\textwidth]{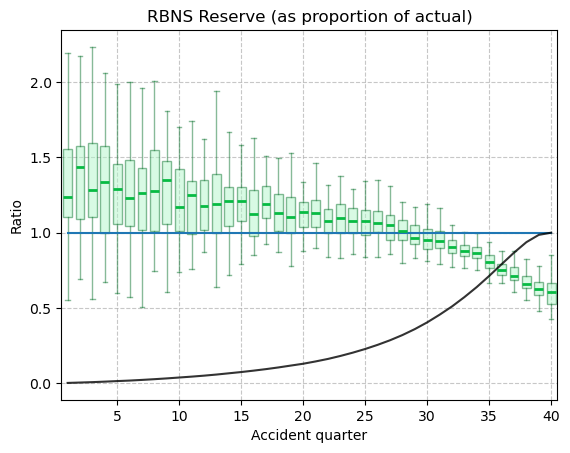}
        \label{fig:FNNplus_OCLs_acc}
    }
    \subfloat[LSTM]{
        \includegraphics[width=0.275\textwidth]{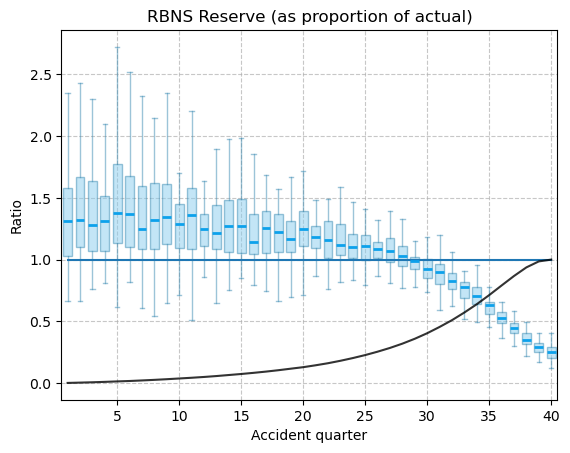}
        \label{fig:LSTM_OCLs_acc}
    }
    \\
    \subfloat[FNN]{
        \includegraphics[width=0.275\textwidth]{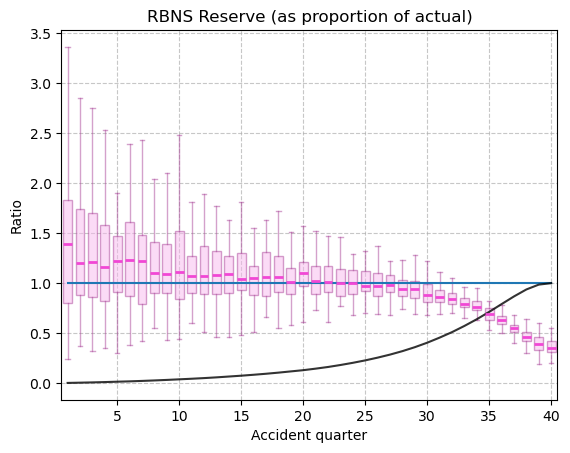}
        \label{fig:FNN_OCLs_acc}
    }
    \subfloat[Case Estimates]{
        \includegraphics[width=0.275\textwidth]{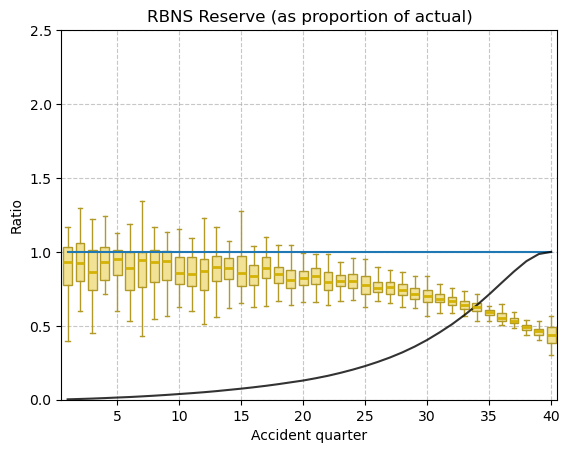}
        \label{fig:CE_OCLs_acc}
    }
    
    \caption{Boxplots showing the aggregate predictions of outstanding claim amounts (expressed as a proportion of the true amounts outstanding) at the valuation date, subset by accident quarter. The black curves represent the proportion of the actual amounts outstanding related to predictions made on or before quarter q.}
    \label{fig:OCLs_acc}

\end{figure}

\begin{figure}[H]
    \centering
    \subfloat[LSTM+]{
        \includegraphics[width=0.275\textwidth]{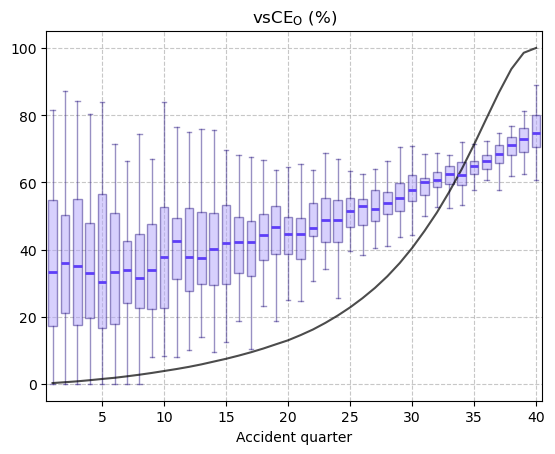}
        \label{fig:LSTMplus_vsCE_OCL_acc}
    }
    \subfloat[FNN+]{
        \includegraphics[width=0.275\textwidth]{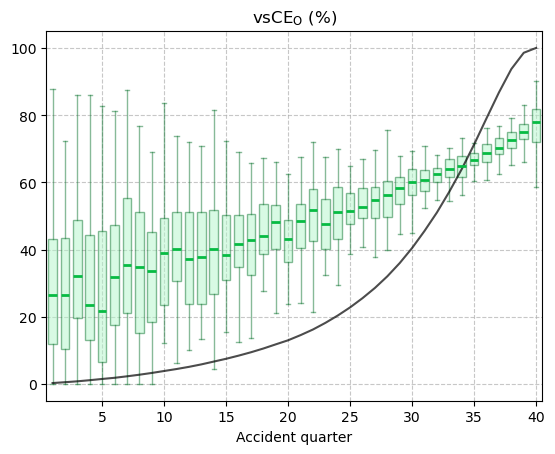}
        \label{fig:FNNplus_vsCE_OCL_acc}
    }
    \\
    \subfloat[LSTM]{
        \includegraphics[width=0.275\textwidth]{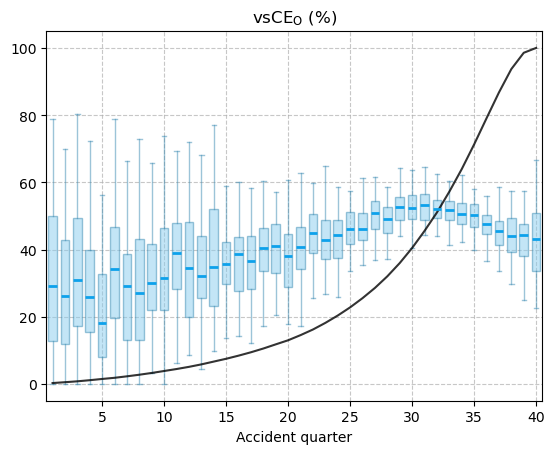}
        \label{fig:LSTM_vsCE_OCL_acc}
    }
    \subfloat[FNN]{
        \includegraphics[width=0.275\textwidth]{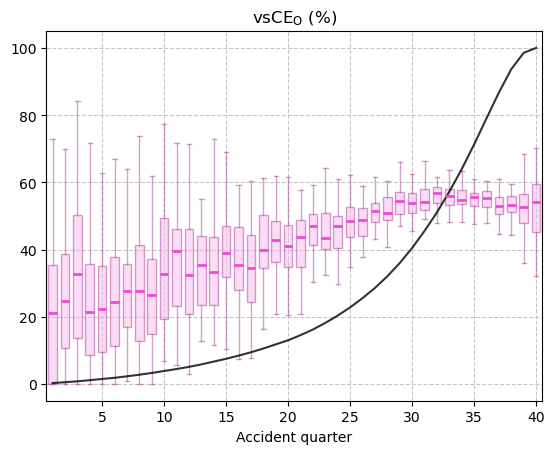}
        \label{fig:FNN_vsCE_OCL_acc}
    }
    
    \caption{Boxplots showing the $\text{vsCE}_\text{O}$ at the valuation date, subset by accident quarter. The black curves represent the proportion of the actual amounts outstanding related to predictions made on or before quarter q.}
    \label{fig:vsCE_OCL_acc}
\end{figure}

\section{Hyperparameter tuning} \label{app:hyperparameter}

\begin{table}[!htb]
    \centering
    \begin{tabular}{@{}ccccc@{}}
    \toprule
    Hyperparameter       & LSTM+                & LSTM                 & FNN+                  & FNN                  \\ \midrule
    Layers               & 2, \textbf{3}, 4     & \textbf{2}, 3, 4     & 2, 3, \textbf{4}      & 2, \textbf{3}, 4     \\
    Units per layer      & \textbf{16}, 32      & 16, \textbf{32}      & \textbf{16}, 32       & 16, \textbf{32}      \\
    Activation function  & ReLU                 & ReLU                 & ReLU                  & ReLU                 \\
    Optimiser            & AdamW                & AdamW                & AdamW                 & AdamW                \\
    Learning rate        & \textbf{0.01}, 0.1   & \textbf{0.01}, 0.1   & \textbf{0.01}, 0.1    & 0.01, \textbf{0.1} \\
    Maximum epochs       & 200                  & 200                  & 200                   & 200                  \\
    Patience             & 5                    & 5                    & 5                     & 5                    \\
    Batch size           & 256, \textbf{512}    & \textbf{256}, 512    & 256, \textbf{512}     & 256, \textbf{512}    \\
    Dropout rate         & 0                    & 0                    & 0                     & 0                    \\ 
    Normalisation layers &  batch and layer     & batch and layer      & batch                 & batch                \\
    Loss function        &  MSE                 & MSE                  & MSE                   & MSE                  \\  \bottomrule
    \end{tabular}
    \caption{Hyperparameters and tested values for each model. Where more than one value is tested for a given hyperparameter, the values selected for the models and datasets presented in sections \ref{q1}, \ref{q2} and \ref{q3} are bolded}.
    \label{fig:hyperparameters}
\end{table}

\subsection*{Number of layers}

This hyperparameter refers to slightly different things depending on the model. For the LSTM and LSTM+, it refers to the number of LSTM layers before combining with the static inputs in two feed-forward layers. For the FNN and FNN+, it refers to the total number of feed-forward layers. We decided to test all models with 2-4 layers.

\subsection*{Number of units per layer}

For the LSTM and LSTM+, this refers to the number of units in each LSTM layer. We tested 16 and 32 units. For the FNN and FNN+, we tested 16 and 32 units per feed-forward layer.

\subsection*{Activation functions}

Unlike the previous two hyperparameters, the choice of activation function applies solely to the feed-forward layers for all models. For the final feed-forward layer, a linear activation is used. For all other feed-forward layers, ReLU is used. No other activation functions have been tested in this layer.

\subsection*{Optimiser and learning rate}

The optimiser refers to the algorithm used to update the parameters of the model during training, with the learning rate specifying how dramatic these updates should be. We tested the AdamW optimiser \citep*{LoHu17} with learning rates of 0.01 and 0.1.

\subsection*{Epochs and early stopping}

An epoch refers to the process of feeding the entire training set into the model (forward pass) and using the errors to update the parameters of the model (backward pass). The way in which the parameters are updated depends on the optimisation algorithm, as mentioned above.

As the number of epochs increases, the model should improve its performance on the training set. However, this is not always ideal due to the risk of overfitting to the training data. Rather, it is important for the model to maximise its performance on unseen data as opposed to memorising the data it has already seen during training. One way to mitigate this risk is through early stopping. 

As the name suggests, this can terminate the training process before the final epoch is reached. At the end of every epoch, the validation set is passed forward through the network and its results are stored. If there are no improvements on the validation set after a given number of epochs, then training will end. This given number of epochs is called `patience'.

After some initial testing, we decided to use a maximum of 200 epochs with a patience of 5 epochs. Once training stops, the weights are `restored' to those that produced the smallest observed loss on the validation set.

\subsection*{Batch size}

Within each epoch, the training data is fed into the model in batches. That is, instead of performing a backward pass after seeing every observation in the training data, the model will instead update its parameters after seeing each subset of the training data.

Determining an appropriate batch size requires a trade-off between computational speed and convergence speed, as well as memory usage. We considered both 256 and 512 as our batch sizes for all models.

\subsection*{Dropout rate}

Dropout \citep{SrHiKrSuSa14} is a form of regularisation that prevents the model from overfitting to the training set. During each forward pass in training, a random number of units in each layer will be set to 0. The probability of a given unit dropping out is equal to the dropout rate. In theory, this prevents the network from becoming too reliant on a single unit and hence improves the network's ability to generalise.

Non-zero dropout rates were initially tested, however the results were not compelling. Consequently, a dropout rate of 0 was adopted for all models.

\subsection*{Normalisation}

Layer normalisation \citep*{BaKiHi16} is applied after each LSTM layer, and is hence only applicable to the LSTM and LSTM+. Batch normalisation \citep*{IoSz15}, however, is present in all of our models. It is applied immediately before the activation function in each feed-forward layer, except for the final feed-forward layer where no normalisation is used.

\section{Results with na\"{\i}ve split} \label{sec:naive results}

In addition to the results from Section \ref{q1}, we also present the same results under a more `na\"{\i}ve' train-test split. Instead of splitting the training, validation and test sets by the finalisation time of each claim and then randomly shifting a section of the validation set into the training, here we use a random split by claim ID across the training, validation and test sets. A 60-20-20 split was used for the proportion of claims in the training, validation and test sets, respectively. 

This split is considered to be na\"{\i}ve as on the surface, it is a common method used to partition the data in such a way that can be used to train a model and evaluate its performance on unseen data. However, it introduces a few issues. Firstly, since the training and validation sets will now contain some claims that would be open as at the valuation date, any reserve that is constructed as of the valuation date will either be incomplete if it only comprises claims from the test set, or it would be biased from the open claims in the training and validation sets. In this section, we choose to present the results based on claims that only appear in the test set and are open at the same valuation date used under our primary splitting method.

\begin{table}[!htb]
    \centering
    \begin{tabular}{@{}crrrrc@{}}
    \toprule
    Model     & ResErr (\%) & MALE          & MSLE           \\ \midrule
    Case Estimates & -33.7 (2.3)    & 0.832 (0.036) & 2.071 (0.139)   \\
    LSTM+     & 4.3 (3.6)      & 0.444 (0.034) & 0.708 (0.333)   \\ 
    FNN+      & 5.0 (3.7)      & 0.426 (0.024) & 0.400 (0.069)   \\ \bottomrule
    \end{tabular}
    \caption{Reserve error, MALE and MSLE metrics at the valuation date across 50 datasets for the LSTM+ and FNN+ under the na\"{\i}ve split. Values are expressed as mean (standard deviation).}
    \label{results table q1 Random}
\end{table}

\begin{table}[!htb]
    \centering
    \begin{tabular}{@{}crrrrc@{}}
    \toprule
    Model           & Case Estimates  & FNN+       \\ \midrule
    LSTM+           & 61.8 (2.8)      & 51.1 (5.4) \\ 
    FNN+            & 61.0 (3.1)      & -          \\ \bottomrule
    \end{tabular}
    \caption{`vs table' for the LSTM+ and FNN+ under the na\"{\i}ve split. Row $i$, column $j$ displays the `M$i$vsM$j$' metric. Values are expressed as mean (standard deviation).}
    \label{vsCE table q1 Random}
\end{table}

\begin{table}[!htb]
    \centering
    \begin{tabular}{@{}crrrrc@{}}
    \toprule
                   & $\lvert \text{ResErr} \rvert$ & MALE  & MSLE & $\text{vsCE}_\text{O}$ \\ \midrule
    Test statistic & 556                           & 328   & 52   & 448   \\
    $p$-value      & 0.4376 & 0.0024 & $7.2 \times 10^{-11}$ & 0.0678     \\ 
    Win count      & 29                            & 18    & 5    & 31   \\\bottomrule
    \end{tabular}
    \caption{Wilcoxon two-sided test statistic and $p$-value comparing the LSTM+ and FNN+ under the na\"{\i}ve split for each evaluation metric. The win count represents the number of datasets (out of 50) where the LSTM+ produced a smaller value than the FNN+ for the absolute reserve error, MALE and MSLE metrics, or a larger value for the $\text{vsCE}_\text{O}$ metric.}
    \label{wilcoxon q1 random}
\end{table}

\begin{figure}[!htb]
    \centering
    \subfloat[MALE]{
        \includegraphics[height=4.5cm, keepaspectratio]{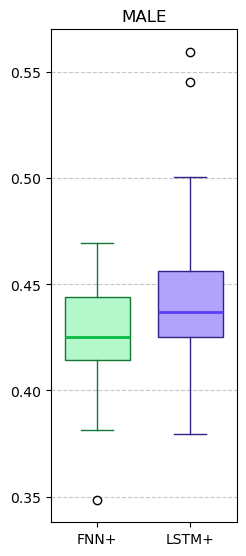}
        \label{fig:LSTMplus_FNNplus_MALE_val_Random}
    }
    \subfloat[MSLE]{
        \includegraphics[height=4.5cm, keepaspectratio]{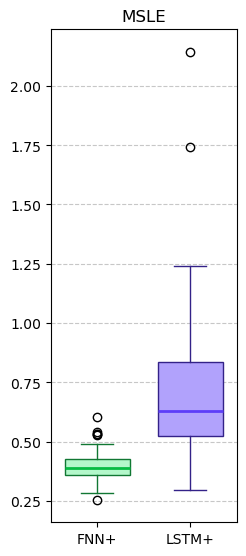}
        \label{fig:LSTMplus_FNNplus_MSLE_val_Random}
    }
    \subfloat[ResErr]{
        \includegraphics[height=4.5cm, keepaspectratio]{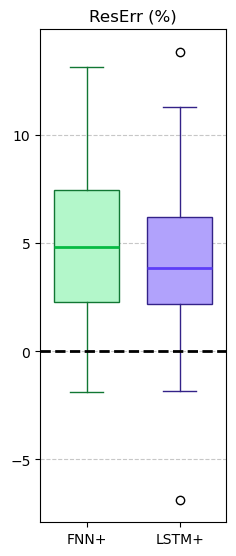}
        \label{fig:LSTMplus_FNNplus_OCLs_val_Random}
    }
    \subfloat[$\text{vsCE}_\text{O}$]{
        \includegraphics[height=4.5cm, keepaspectratio]{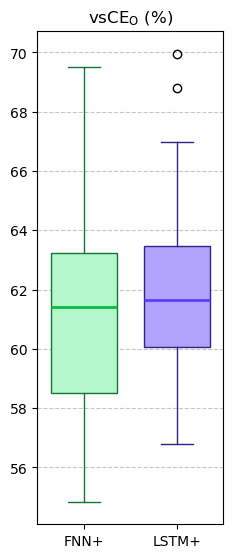}
        \label{fig:LSTMplus_FNNplus_vsCE_OCL_val_Random}
    }

    \caption{Boxplots containing MALE, MSLE, reserve error and $\text{vsCE}_\text{O}$ metrics for the LSTM+ and FNN+ under the na\"{\i}ve split at the valuation date. Smaller values for \protect\subref{fig:LSTMplus_FNNplus_MALE_val_Random} and \protect\subref{fig:LSTMplus_FNNplus_MSLE_val_Random}, values closer to 0 for \protect\subref{fig:LSTMplus_FNNplus_OCLs_val_Random} and larger values for \protect\subref{fig:LSTMplus_FNNplus_vsCE_OCL_val_Random} indicate stronger performance.}
    \label{fig:boxplots-val LSTMplus_FNNplus Random}
\end{figure}

\noindent Tables \ref{results table q1 Random} and \ref{vsCE table q1 Random} display the results from the LSTM+ and FNN+ models shown earlier, after retuning and training on the datasets under the na\"{\i}ve split. Compared to Tables \ref{results table q1} and \ref{vsCE table q1}, the mean and standard deviation of reserve errors in absolute terms have reduced. The mean MALE and MSLE are also significantly lower under the na\"{\i}ve split, which together with the smaller reserve errors suggest an `improvement' at both an aggregate and an individual claim level, or rather an overstatement of their performances. This is again reinforced by comparing Figure \ref{fig:boxplots-val LSTMplus_FNNplus Random} (the boxplots of MALE, MSLE, reserve error and $\text{vsCE}_\text{O}$ under the na\"{\i}ve split) with Figure \ref{fig:boxplots-val LSTMplus_FNNplus} (the same boxplots under the originally proposed split).

\begin{figure}[!htb]
    \centering
    \subfloat[LSTM+]{
        \includegraphics[height=4.5cm, keepaspectratio]{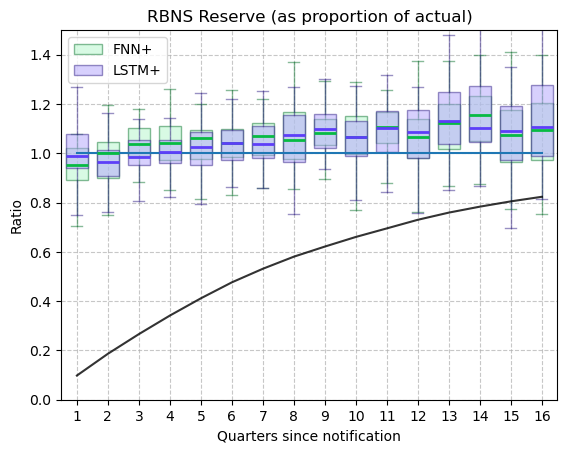}
        \label{fig:LSTMplus_FNNplus_OCLs_dev_Random}
    }
    \subfloat[FNN+]{
        \includegraphics[height=4.5cm, keepaspectratio]{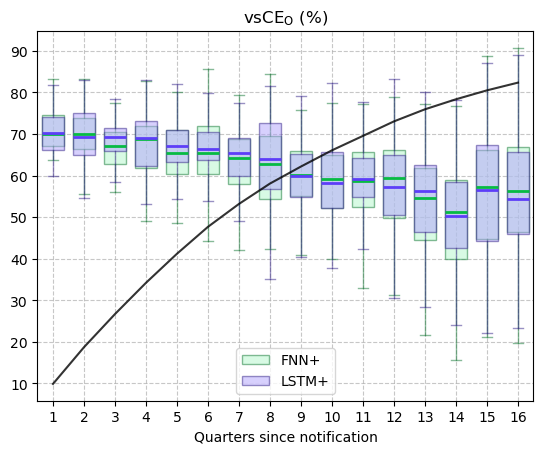}
        \label{fig:LSTMplus_FNNplus_vsCE_OCL_dev_Random}
    }
    \caption{Boxplots showing the aggregate predictions of outstanding claim amounts (expressed as a proportion of the true amounts outstanding) and $\text{vsCE}_\text{O}$ at the valuation date for models under the na\"{\i}ve split, subset by quarters since notification. The black curves represent the proportion of the actual amounts outstanding related to predictions made on or before quarter q. Stronger performance is indicated by values closer to 1 for \protect\subref{fig:LSTMplus_FNNplus_OCLs_dev_Random} and larger values for \protect\subref{fig:LSTMplus_FNNplus_vsCE_OCL_dev_Random}.}
    \label{fig: LSTMplus_FNNplus Random}
\end{figure}

While the aforementioned results appear more favourable for both models under the na\"{\i}ve split, this point is most clearly shown in the aggregate forecasts by quarter since notification, as shown in Figure \ref{fig: LSTMplus_FNNplus Random}. Here, both models show a remarkably tight forecast across all stages of development. It can be argued that the models perform better in early periods of development, however this is partially driven by the bias correction. Regardless, these models show a significant improvement in quarters 1-5, which accounts for almost 50\% of the true outstanding claim amounts at the valuation date. Therefore, an unwitting use of the na\"{\i}ve split may lead to conclusions that the models are able to produce reasonable predictions in aggregate across all development quarters, when in reality the models tend to struggle in early periods of development.

Even though performance has been artificially enhanced for both the LSTM+ and FNN+, the research question initially asked for a comparison between the two models. Table \ref{wilcoxon q1 random}, when compared to Table \ref{wilcoxon q1}, tells a similar story. The MALE and MSLE continue to be significantly smaller for the FNN+, while the test results for the absolute reserve error and the $\text{vsCE}_\text{O}$ are no longer significant at the 5\% level of significance. Therefore, similar conclusions can be drawn to those from Section \ref{q1} in that the FNN+ appears to slightly outperform the LSTM+ at an individual claim level, while the results at an aggregate level are mixed.

Ultimately, the relative performance of each model is quite similar under both splitting methods. This suggests that while in this case the answer to the research question is not significantly altered by the use of a na\"{\i}ve split, it does still present an overly optimistic view of how the models will ultimately perform in practice.

\end{document}